\documentclass[onecolumn]{IEEEtran}
\usepackage{mathpazo}
\usepackage{times}
\usepackage{amsmath}
\usepackage{amsfonts}
\usepackage{latexsym}
\usepackage{amssymb}
\usepackage[noadjust]{cite}
\usepackage[ruled,lined]{algorithm2e}
\usepackage{upref}
\usepackage{theorem}
\usepackage{graphicx}
\usepackage{overpic}

\usepackage[usenames]{color}
 \usepackage {tikz}
\usetikzlibrary {positioning}

\usepackage{todonotes}
\setuptodonotes{fancyline}
\usepackage{mathtools}
\usepackage{xparse}
\DeclarePairedDelimiterX{\set}[1]{\{}{\}}{\setargs{#1}}
\DeclarePairedDelimiterX{\bset}[1]{[}{]}{\setargs{#1}}
\NewDocumentCommand{\setargs}{>{\SplitArgument{1}{;}}m}
{\setargsaux#1}
\NewDocumentCommand{\setargsaux}{mm}
{\IfNoValueTF{#2}{#1} {#1\,\delimsize|\,\mathopen{}#2}}

\DeclarePairedDelimiter\abs{\lvert}{\rvert}
\DeclarePairedDelimiter\norm{\lVert}{\rVert}
\DeclarePairedDelimiter\ceil{\lceil}{\rceil}

\DeclarePairedDelimiter\parenv{\lparen}{\rparen}

\theoremstyle{plain}
\newtheorem{theorem}{Theorem}
\newtheorem{lemma}[theorem]{Lemma}

\newtheorem{corollary}[theorem]{Corollary}
\newtheorem{definition}[theorem]{Definition}
\newtheorem{example}[theorem]{Example}

\newtheorem{construction}{Construction}

\newcommand{\cF}{\mathcal{F}}

\newcommand{\cP}{\mathcal{P}}
\newcommand{\cQ}{\mathcal{Q}}

\newcommand{\cT}{\mathcal{T}}

\renewcommand{\leq}{\leqslant}

\renewcommand{\geq}{\geqslant}

\newcommand{\N}{\mathbb{N}}
\newcommand{\R}{\mathbb{R}}
\newcommand{\Z}{\mathbb{Z}}
\newcommand{\E}{\mathbb{E}}

\newcommand{\ccap}{\mathsf{cap}}

\DeclareMathOperator{\factor}{Factor}
\DeclareMathOperator{\irr}{Irr}
\DeclareMathOperator{\fr}{fr}

\newcommand{\eqdef}{\triangleq}

\newcommand{\com}[1]{\overline{#1}}
\newcommand{\der}{\Longrightarrow}
\newcommand{\Trc}{T^{\mathrm{rc}}}
\newcommand{\cTrc}{\cT^{\mathrm{rc}}}
\newcommand{\Src}{S^{\mathrm{rc}}}

\newcommand{\T}{\intercal}
\newcommand{\rc}{{\mathrm{rc}}}

\begin{document}


\title{On the Reverse-Complement\\ String-Duplication System}

\author{
Eyar Ben-Tolila and Moshe~Schwartz,~\IEEEmembership{Senior Member,~IEEE}%
\thanks{Eyar Ben-Tolila is with the School
   of Electrical and Computer Engineering, Ben-Gurion University of the Negev,
   Beer Sheva 8410501, Israel
   (e-mail: eyarb@post.bgu.ac.il).}%
\thanks{Moshe Schwartz is with the School
   of Electrical and Computer Engineering, Ben-Gurion University of the Negev,
   Beer Sheva 8410501, Israel
   (e-mail: schwartz@ee.bgu.ac.il).}%
\thanks{This work was supported in part by the Israel Science Foundation (ISF) under grant No.~270/18.}
}

\maketitle

\begin{abstract}
Motivated by DNA storage in living organisms, and by known biological mutation processes, we study the reverse-complement string-duplication system. We fully classify the conditions under which the system has full expressiveness, for all alphabets and all fixed duplication lengths. We then focus on binary systems with duplication length $2$ and prove that they have full capacity, yet surprisingly, have zero entropy-rate. Finally, by using binary single burst-insertion correcting codes, we construct codes that correct a single reverse-complement duplication of odd length, over any alphabet. The redundancy (in bits) of the constructed code does not depend on the alphabet size.
\end{abstract}


\section{Introduction}

\IEEEPARstart{D}{NA} is a very appealing medium for storing digital information, whose rate of creation is growing exponentially in recent years, causing an expanding gap between information production and data storage capabilities. Compared with current storage technologies, DNA offers densities that are higher in orders of magnitude, and thus can potentially serve as an extremely efficient storage system. Already, a density of $2.15\cdot 10^{17}$ bytes per gram of DNA molecules has been demonstrated~\cite{ErlZie17}, whereas the densest commercially available option~\cite{nimbus} is capable of storing only $1.86\cdot 10^{11}$ bytes per gram of hardware.

Data can be stored in DNA, in vitro or in vivo. While the former will likely provide a higher density, the latter has many advantages. First, in-vivo DNA storage can serve as a protected medium for storing large amounts of data in a compact format for long periods of time~\cite{Bal13,WonWonFoo03}. An additional advantage is that data can be disguised as part of the organisms’ original DNA, thus providing an added layer of secrecy~\cite{CleRisBan99}. Finally, in-vivo DNA storage has further applications such as watermarking genetically modified organisms, and enabling synthetic biology methods~\cite{Micetal12,ShiNivMacChu17,HeiBar07}.

However, storing information in living organisms introduces new types of errors. Among these new error types we find duplication errors, which are motivated by a class of mutations that are common in most organisms and lead to an abundance of repeated sequences in their genomes. Some examples are transposon-driven repeats~\cite{Lanetal01}, and tandem repeats, which are believed to be caused by slipped-strand mispairings~\cite{Lev87}. These mutation
processes take a substring of the DNA and insert a copy of it into the original string. The copy can be inserted anywhere in the string, and might even be reversed and complemented during the insertion~\cite{Ger08}. A formal mathematical model for studying these kinds of mutation processes is the notion of string-duplication systems~\cite{FarSchBru16}. In such systems, a seed string evolves over time by successive applications of duplication functions.

Previous works studied several properties of these models, among which we mention capacity, entropy rate, and expressiveness (the ability to create any possible target substring). It was shown in~\cite{FarSchBru16} that the system with end duplication, which copies a substring of a fixed length $k$ to the end of the string, has full expressiveness and full capacity. In contrast, the system with a fixed length tandem duplication, where the copy is inserted next to its original position, has zero capacity and is never fully expressive, but when the length of the tandem duplication is only lower-bounded, full expressiveness and positive capacity are obtained.  In~\cite{EliFarSchBru19}, the exact entropy rate of both end- and tandem-duplication systems were found, in the case where the duplication size is $k=1$ and the alphabet is binary. Furthermore, a noisy scenario was investigated where the duplicated bit has a probability of being complemented. In that case, the exact entropy rate of end-duplication systems was found, as well as bounds on the entropy rate of tandem-duplication systems. In~\cite{LouSchBruFar20}, using stochastic approximation methods, upper bounds were found on the the entropy rate of tandem duplication with a probability of substitutions.

To protect against these duplications, error-correcting codes for duplication channels were studied as well. A construction correcting any number of tandem duplications of a fixed length $k$ was found in~\cite{JaiFarSchBru17a}. Other codes that correct a prescribed number of duplications were constructed in~\cite{LenWacYaa19,ZerEsmGul19,KovTan18,Kov19}, addressing several duplication types. Additional codes that are capable of correcting a mixture of tandem duplications and substitutions or edits were described in~\cite{TanYehSchFar20,TanFar21,TanFar22}. Finally, some Levenshtein-reconstruction problems for duplications were studied in~\cite{YehSch20,YehSch21}.

In this work we study reverse-complement duplication, in which the duplicated copy is reversed and complemented before being inserted next to its original location. This duplication process has been observed in the genomes of many living organisms~\cite{ButGilSte02,DieTanBerYaoTap10}. Previous works on this model were severely limited. In~\cite{FarSchBru16} the duplicated copy was reversed but not
complemented. It was shown there that the system with a fixed-length duplication has full expressiveness and positive capacity. In~\cite{EliFarSchBru19}, reverse-complement duplication was studied, but only for a duplication of length $k=1$. Upper and lower bounds on the entropy rate of the system were found there, but a gap between them remains. Therefore, many aspects of this system are yet to be studied.

We make the following contributions in this paper. First, we fully classify the exact conditions under which the reverse-complement string-duplication system has full expressiveness, for any alphabet and any fixed duplication length. Next, we prove that the binary system with duplication length $2$ has full capacity, by carefully characterizing the irreducible strings. We then continue to a probabilistic setting, and show that the same system has zero entropy rate, which is surprising considering the fact that it has full capacity. Finally, we construct error-correcting codes that are capable of fixing a single duplication of odd length, over any alphabet. The coding scheme is built on burst-insertion-correcting codes, and interestingly, has redundancy (in bits) that does not depend on the alphabet size.

The paper is organized as follows. We begin in Section~\ref{sec:prelim}, by introducing basic concepts and notation that will be used throughout the paper. Then, in Section~\ref{sec:express}, we completely classify the expressive power of the reverse-complement string-duplication system. We continue in Section~\ref{sec:capentropy}, by studying the capacity and entropy rate of these systems. Error-correcting codes are constructed in Section~\ref{sec:ecc}. We conclude in Section~\ref{sec:conc} by summarizing our results and discussing open problems.

\section{Preliminaries}
\label{sec:prelim}

Throughout this paper we let $\Sigma$ denote some finite alphabet, the elements of which are called letters. A sequence of letters is a string, e.g., $u=u_0 u_1 \dots u_{n-1}$, where $u_i\in\Sigma$ for all $i$. In this case we say the length of $u$ is $n$, and we denote it by $\abs{u}=n$. The set of all strings of length $n$ is denoted by $\Sigma^n$. The unique string of length $0$ is denoted by $\varepsilon$, i.e., $\Sigma^0=\set{\varepsilon}$. We then define $\Sigma^*$ to be the set of all strings of finite length, i.e., $\Sigma^*\eqdef \bigcup_{i\geq 0}\Sigma^i$. Similarly, we define $\Sigma^+\eqdef \Sigma^*\setminus\set{\varepsilon}$ to be the set of all non-empty strings of finite length.

Given two strings, $u\in\Sigma^n$ and $v\in\Sigma^m$, we write $uv$ to denote the string of length $n+m$ formed by the concatenation of $u$ and $v$. For any $\ell\in\N$, we write $u^\ell$ to denote the string of length $\ell n$ formed by concatenating $\ell$ copies of $u$. We define $u^0=\varepsilon$. We further define
\[ u^*\eqdef \set*{u^\ell ; \ell\geq 0}, \qquad u^+\eqdef\set*{u^\ell ; \ell\geq 1}.\]
We say that $y\in\Sigma^*$ is a \emph{prefix} of $w\in\Sigma^*$ if there exists $z\in\Sigma^*$ such that $w=yz$. Similarly, $y$ is a \emph{suffix} of $w$ if there exists $x\in\Sigma^*$ such that $w=xy$. Also, $y$ is a \emph{factor} (or \emph{substring}) of $w$ if there exist $x,z\in\Sigma^*$ such that $w=xyz$. We shall say that $y$ is a $k$-prefix (respectively, $k$-suffix, $k$-factor) of $w$, if it is a prefix (respectively, suffix, factor) of $w$, and $\abs{y}=k$. If $S\subseteq\Sigma^*$ is a set of strings, its factor set is then defined as
\[ \factor(S) \eqdef \set*{ v\in\Sigma^* ; \exists u,w\in\Sigma^* \text{ s.t. } uvw\in S}.\]

We assume throughout the paper, that a \emph{complement} operation is defined over $\Sigma$. More precisely, we assume that for every letter $a\in\Sigma$ there exists a unique letter $b\in\Sigma$, $b\neq a$, which is the complement of $a$. We denote this by $\com{a}=b$. This is a symmetric relation, namely, $\com{b}=a$. As a consequence, $\abs{\Sigma}$ must be even. We extend the complement operation to strings in the natural way. That is, if $u=u_0 u_1 \dots u_{n-1}\in\Sigma^n$, then $\com{u}=\com{u_0}\, \com{u_1} \dots \com{u_{n-1}}$ is the letter-wise complement string.

Another useful notation we introduce is that of string reversal. Let $u=u_0 u_1 \dots u_{n-1}\in\Sigma^n$ be a string of length $n$. The \emph{reversal} of $u$ is denoted by $u^R\eqdef u_{n-1} u_{n-2} \dots u_0$. Obviously, the reversal and complement operations are independent, and so $\com{u}^R=\com{u^R}$.

We now turn to describe the string-duplication framework. We follow the ideas and notation as described in~\cite{FarSchBru16}. A \emph{string-duplication rule} is simply a function $T:\Sigma^*\to\Sigma^*$. Following the motivation for this framework, such rules describe, or are inspired by, biological processes that create duplications during DNA replication. The rules that were studied in~\cite{FarSchBru16} were tandem duplication, end duplication, interspersed duplication, and palindromic duplication. Let $\cT\subseteq \Sigma^{*\Sigma^*}$ denote a set of such duplication rules. For a string $u\in\Sigma^*$, we say that $v$ is an $\ell$-descendant of $u$, if there exist $T_1,\dots,T_\ell\in \cT$, not necessarily distinct, such that $v=T_\ell(T_{\ell-1}(\dots T_1(u)\dots))$, and we denote it by $u\der^\ell v$. If $\ell=1$ we just write $u\der v$. Additionally, for $\ell=0$ we only have $u\der^0 u$. The set of all $\ell$-descendants of $u$ is the set
\[ D^\ell(u) \eqdef \set*{ v\in\Sigma^* ; u\der^\ell v}.\]
The \emph{descendant cone} of $u$ is then defined as all the strings which may be derived from $u$ following a finite number of duplication rules, namely,
\[ D^*(u) \eqdef \bigcup_{\ell\geq 0}D^\ell(u).\]
We can then write $u\der^* v$ if and only if $v\in D^*(u)$.

A \emph{string-duplication system}, $S(\Sigma,s,\cT)$, where $\Sigma$ is a finite alphabet, $s\in\Sigma^*$ is the \emph{seed string}, and $\cT\subseteq \Sigma^{*\Sigma^*}$ is a set of duplication rules, is defined as the descendant cone of $s$, that is
\[ S(\Sigma,s,\cT) \eqdef D^*(s).\]
Thus, $S$ contains all the strings that may result from $s$ after applying a finite number of arbitrary duplication rules from $\cT$.

The dual of the descendant cone is the \emph{ancestor cone}. Here, the ancestor cone of $u\in\Sigma^*$ is defined as
\[ A^*(u) \eqdef \set*{v\in\Sigma^* ; u\in D^*(v)},\]
namely, all the strings of which $u$ is a descendant. We note that $u\in A^*(u)$ always. However, if $A^*(u)=\set{u}$, then we say that $u$ is \emph{irreducible}.

Fix some string-duplication system $S=S(\Sigma,s,\cT)$. Depending on the application, one can view $S$ in two ways: either as a generative model which describes what possible strings may be derived from $s$ (e.g., see~\cite{FarSchBru16,JaiFarBru17,AloBruFarJai17,CheChrKiaNgu19,EliFarSchBru19,LouSchBruFar20}), or as a channel which describes what corrupted versions of the transmitted $s$ may be received (e.g., see~\cite{JaiFarSchBru17a,KovTan18,YehSch20,LenWacYaa19,Kov19,ZerEsmGul19,TanYehSchFar20,TanFar21,YehSch21}). While the latter view calls for the construction of suitably tailored error-correcting codes, the former inspires the following properties of $S$ to be studied.

The first property of $S$ is full expressiveness. We say that $S$ is \emph{fully expressive} if for any $v\in\Sigma^*$ there exist $u,w\in\Sigma^*$ (that may depend on $v$) such that $uvw\in S$. Thus, in a fully expressive system, any finite string appears as a factor of one of the strings in the system (see~\cite{FarSchBru16,JaiFarBru17}). 

The second property of $S$ is its capacity. We define the \emph{capacity} of $S$ to be
\[ \ccap(S) \eqdef \limsup_{n\to\infty}\frac{1}{n}\log_2\abs*{S\cap \Sigma^n}.\]
Intuitively, the capacity of $S$ measures the exponential growth rate of the descendant cone of the seed string. One can trivially see that $\ccap(S)\leq \log_2\abs{\Sigma}$, and if equality holds we say that $S$ has full capacity. Additionally, as mentioned in~\cite{FarSchBru16}, if $S$ has full capacity then it must be fully expressive.

The last property of $S$ is its entropy rate (see~\cite{EliFarSchBru19}). Unlike the deterministic nature of the previous two properties, here we describe a stochastic process. Denote $S(0)=s$, the seed string. Then, at each step $i=1,2,\dots$, using some probability distribution (that may depend on $i$) over $\cT$, we apply a randomly chosen duplication rule from $\cT$ to $S(i-1)$, thus obtaining $S(i)$. Hence, $S(n)$, $n\in\N$, are all random variables. We can then define the entropy of $S(n)$ as
\[ H(S(n))\eqdef -\sum_{w\in\Sigma^*}\Pr(S(n)=w)\log_2\Pr(S(n)=w).\]
With this, the \emph{entropy rate} of the random process $S$ is defined as
\[ h(S)\eqdef \limsup_{n\to\infty}\frac{1}{n}H(S(n)).\]
Loosely speaking, $h(S)$ measures the amount of information generated by an application of a random duplication rule.

We conclude this section by describing the specific set of duplication rules we shall be studying in this paper. Let $\Sigma$ be a finite alphabet with a complement operation defined on it. The reverse-complement duplication rule, that copies a $k$-factor starting in position $i$ in the given string, is defined for all $x\in\Sigma^*$ as
\[ \Trc_{i,k}(x)\eqdef \begin{cases}
uv\com{v}^R w, & \text{if $x=uvw,\abs{u}=i,\abs{v}=k$,}\\
x, & \text{otherwise.}
\end{cases}\]
We then define the set of duplication rules, for a fixed duplication length $k\in\N$ to be
\[ \cTrc_k\eqdef \set*{\Trc_{i,k}; i\geq 0}.\]
Finally, the \emph{$k$-uniform reverse-complement string-duplication system} is defined as
\[ \Src_k \eqdef S(\Sigma,s,\cTrc_k).\]
We note that in the notation $\Src_k$, the dependence on $\Sigma$ and $s$ is implicit. If this dependence may not be easily inferred from the context, or if a need for emphasis arises, we shall write this dependence explicitly.

\begin{example}
Consider the alphabet $\Sigma=\Z_4=\set{0,1,2,3}$ with complement pairs defined by $\com{0}=1$ and $\com{2}=3$. Assume that the duplication length is $k=2$. Then,
\[ 0123 \der 012\underline{30}3 \der 0123\underline{23}03,\]
where the duplicated factor is underlined. Thus, $01232303\in \Src_2=S(\Sigma,0123,\cTrc_2)$.
\end{example}

We would like to note that the reverse-complement string-duplication system is indeed biologically motivated. DNA sequences are strings of bases (or nucleotides). Since there are four possible bases, adenine (A), thymine (T), guanine (G), and cytosine (C), we can think of DNA sequences as strings over $\Sigma=\set{A,C,G,T}$. The bases form two complementary pairs, $\com{A}=T$, and $\com{C}=G$. A reverse complement of a section of the DNA molecule might be inserted immediately following the said section, thus creating a \emph{palindromic duplication} (e.g., see~\cite{ButGilSte02,DieTanBerYaoTap10}). The reverse-complement string-duplication system models this phenomenon. We mention in passing that~\cite{FarSchBru16} misused the term ``palindromic duplication'' to describe the insertion of a reversed, but not complemented, copy of a section of a DNA molecule. To avoid confusion with~\cite{FarSchBru16} and papers that cite it, we use the term ``reverse-complement duplication''.

\section{Expressiveness}
\label{sec:express}

In this section we study the expressiveness of the $k$-uniform reverse-complement string-duplication system, $\Src_k$. We show a simple sufficient and necessary condition on the seed string that implies the system is fully expressive when $k\geq 2$. The special case of $k=1$ is also fully characterized.

We first study the case of $k\geq 2$. For any string $s\in\Sigma^*$, and any non-negative integer $i$, we say $a\in\Sigma$ is the \emph{$i$th letter from the end of $s$} if $s=uav$, with $u\in\Sigma^*$ and $v\in\Sigma^i$. The main technical ingredient in proving the expressiveness of $\Src_k$ is the following lemma, which shows that we can push letters towards the end of the string in a controlled manner.

\begin{lemma}
\label{lemma:shortenL}
Let $k\in\N$, $k\geq 2$, $s\in\Sigma^*$, $\abs{s}\geq k+1$, and let $i\geq 2$ be an integer. If the $i$th letter from the end of $s$ is $a$, then there exists $s'\in D^2(s)$, such that $a$ is the $(i-2)$nd letter from the end of $s'$, where descendants are obtained using $k$-uniform reverse-complement duplications.
\end{lemma}

\begin{IEEEproof}
By the lemma's conditions, we can write $s=uvbcw$, where $u,w\in\Sigma^*$, $v\in\Sigma^{k-1}$, $b,c\in\Sigma$, $\abs{w}\leq i-2$, and $v$ contains the letter $a$ that is $i$th from the end of $s$. Now,
\[ s=uvbcw \der uvb\underline{\com{b}\com{v}^R} cw \der uv\com{b}\com{v}^R c\underline{\com{c}v}w=s',\]
where for the reader's convenience we underlined the duplicated parts. The claim now follows immediately, since the letter $a$ in $v$ that was $i$th from the end in $s$, is $(i-2)$nd from the end of $s'$.
\end{IEEEproof}

Assume now that $s=s_0 s_1 \dots s_{n-1}$ is a string of length $n$, $s_i\in\Sigma$ for all $i$. For $j=0,1$ we define
\[ \Sigma_j(s)\eqdef \set*{ s_i ; 1\leq i\leq n, i\equiv j\  (\bmod 2)}.\]
Thus, $\Sigma_0(s)$ (respectively, $\Sigma_1(s)$) is the set of letters of $s$ that appear in even (respectively, odd) positions. 

If $A\subseteq\Sigma$ is a subset of letters, we define
\[ \com{A} \eqdef \set*{\com{a} ; a\in A}.\]
For $s\in\Sigma^*$ the following is a useful definition:
\[ \delta(s)\eqdef \Sigma_0(s)\cup \Sigma_1(\com{s}).\]
We observe that
\begin{equation}
    \label{eq:comdelta}
\delta(\com{s})=\Sigma_0(\com{s}) \cup \Sigma_1(s) = \com{\delta(s)}.
\end{equation}

We are now ready to prove the sufficient and necessary condition for $\Src_k$ to have full expressiveness.

\begin{theorem}
Let $k\in\N$, $k\geq 2$, and let $s\in\Sigma^*$, $\abs{s}\geq k$, be a seed string. Then $\Src_k(s)$ has full expressiveness if and only if:
\begin{enumerate}
    \item $k$ is odd, and $\delta(s)\cup\delta(\com{s})=\Sigma$.
    \item $k$ is even, and $\delta(s)=\Sigma$.
\end{enumerate}
\end{theorem}

\begin{IEEEproof}
We first prove the ``only if'' part of the claim. Consider $k$ odd, and assume that $\delta(s)\cup\delta(\com{s})\neq\Sigma$. It follows that there exists a letter $a\in\Sigma$ such that neither $a$, nor $\com{a}$, appear in $s$. Hence, $a$ and $\com{a}$ cannot appear in any of the descendants of $s$, and $\Src_k(s)$ is not fully expressive. Now consider $k$ even, and assume $\delta(s)\neq \Sigma$. Thus, there exists a letter $a\in\Sigma$ such that $\delta(s)\subseteq\Sigma\setminus\set{a}$, and by~\eqref{eq:comdelta}, also $\delta(\com{s})\subseteq\Sigma\setminus\set{\com{a}}$. Consider now a $k$-factor of $s$ that is being duplicated resulting in $s'$, namely,
\[ s=uvw \der uv\com{v}^R w=s'.\]
Since $k$ is even, we can easily see that
\begin{align*}
\Sigma_0(s') & \subseteq\Sigma_0(s)\cup\Sigma_1(\com{s}) =\delta(s)\subseteq \Sigma\setminus\set{a},\\
\Sigma_1(s') & \subseteq\Sigma_0(\com{s})\cup\Sigma_1(s) =\delta(\com{s})\subseteq \Sigma\setminus\set{\com{a}}.
\end{align*}
Hence, once again we have $\delta(s')\subseteq\Sigma\setminus\set{a}$. By simple induction, this holds not only for $s'$, but for any string in $D^*(s)$. Thus, $aa$ is not a factor of any string in $D^*(s)$, and $\Src_k(s)$ is not fully expressive. 

We move on to prove the ``if'' part. We now contend that for any descendant $u\in D^*(s)$, and any letter $a\in\Sigma$, we can find a descendant $u'\in D^*(u)$ such that the $(2i)$th letter from the end of $u'$ is $a$, for some $i\in\N$ (namely, $a$ is found in an even position from the end of the string). We distinguish between two cases depending on the parity of $k$.

Assume $k$ is odd. The requirement that $\delta(s)\cup\delta(\com{s})=\Sigma$ implies that for any letter $a\in\Sigma$, we have that $a$ or $\com{a}$ appear in $s$. Trivially, any descendant of $u\in D^*(s)$ also satisfies $\delta(u)\cup\delta(\com{u})=\Sigma$, since no letter gets erased in the duplication process. Let $u\in D^*(s)$ be some descendant of $s$. We have the following cases:
\begin{enumerate}
    \item 
    If $u$ contains the letter $a$ as the $(2i)$th letter from the end, $i\in\N$, we are done by setting $u'=u$.
    \item
    Otherwise, if $a$ is the last letter of $u$, we perform two $k$-suffix duplications starting with $u$. The resulting $u'$ has $a$ as the $(2k)$th letter from the end.
    \item
    Otherwise, if $u$ contains $a$ as the $(2i-1)$th letter from the end, perform a single $k$-suffix duplication to obtain $u'$. Now $a$ is the $(2i-1+k)$th letter from the end of $u'$, and $2i-1+k$ is even, since $k$ is odd.
    \item
    Otherwise, $u$ does not contain $a$, but only $\com{a}$. Perform a single duplication on a factor of $u$ that contains $\com{a}$ to obtain $u''$. Now $u''$ contains $a$, and we repeat the arguments from the first three cases to obtain the desired $u'$.
\end{enumerate}

Assume $k$ is even. Imagine the letters of $s$ in even positions are colored red, and those in odd positions green. Furthermore, assume any letters inserted due to duplications are colored blue. Since $k$ is even, one can easily see that in any descendant $u\in D^*(s)$, the red letters remain in even positions, and the green letters remain in odd positions. Hence, $\Sigma=\delta(s)\subseteq\delta(u)$, and so $\delta(u)=\Sigma$. We have the following cases:
\begin{enumerate}
    \item 
    If $u$ contains the letter $a$ as the $(2i)$th letter from the end, $i\in\N$, we are done by setting $u'=u$.
    \item
    Otherwise, if the last letter of $u$ is $a$, perform a single $k$-suffix duplication to obtain $u'$. Then the $k$th letter from the end of $u'$ is $a$.
    \item
    Otherwise, since $\delta(u)=\Sigma$, necessarily $\com{a}$ is the $(2j-1)$st letter from the end, for some $j\in\N$. Perform a single duplication on a factor of $u$ that contains $\com{a}$ to obtain $u'$. Now $u'$ contains $a$ as the $(2i)$th letter from the end, for some $i\in\N\cup\set{0}$. If $i=0$, i.e., $a$ is the last letter of $u'$, perform another $k$-suffix duplication so $a$ is the $k$th letter from the end.
\end{enumerate}

We have therefore proved our auxiliary claim, namely, that for any $u\in D^*(s)$ and any $a\in\Sigma$, there exists $u'\in D^*(u)$ such that $a$ is the $(2i)$th letter from the end of $u'$ for some $i\in\N$. We now claim that there exists $u''\in D^*(u')$ such that $u''$ ends with $a$. This is accomplished by repeated application of Lemma~\ref{lemma:shortenL}, provided $\abs{u'}\geq k+1$. If, $\abs{u'}=k$, take $u'\com{u'}^R u'\in D^2(u')$ and apply Lemma~\ref{lemma:shortenL} repeatedly on it to obtain the desired $u''$. Note that in all cases we apply Lemma~\ref{lemma:shortenL} at least once.

To complete the proof of the ``if'' part, assume we are given a string $a_0 a_1 \dots a_{\ell-1}\in\Sigma^\ell$. We intend to show that there exists a descendant of $s$ whose $\ell$-suffix is $a_0 \dots a_{\ell-1}$. By our previous discussion we can derive from $s$ a string that ends with $a_{\ell-1}$,
\[ s \der^* v a_{\ell-1}.\]
Since at least one application of Lemma~\ref{lemma:shortenL} was used in the process, we necessarily have that $\delta(v)\cup\delta(\com{v})=\Sigma$ if $k$ is odd, and $\delta(v)=\Sigma$ if $k$ is even. We can therefore repeat the argument, starting with $v$, to derive a string that ends with $a_{\ell-2}$. In context of the entire derivation starting from $s$ we obtain,
\[ s \der^* v a_{\ell-1} \der^* v' a_{\ell-2} a_{\ell-1}.\]
By simple induction, we can repeat the process until
\[ s \der^* v'' a_0 a_1 \dots a_{\ell-1},\]
thus completing the proof.
\end{IEEEproof}

When the duplication-window size is $k=1$ we have a different situation. Here, full expressiveness depends solely on whether the alphabet is binary or not.

\begin{theorem}
Let $k=1$, and let $s\in\Sigma^*$, $\abs{s}\geq k$, be a seed string. Then $\Src_k(s)$ has full expressiveness if and only if $\abs{\Sigma} = 2$.
\end{theorem}
\begin{IEEEproof}
In the first direction, assume $\abs{\Sigma}=2$, and w.l.o.g., suppose $\Sigma = \set{0,1}$. It is trivial that, starting with $0$, we can derive any binary string that starts with $01$. First, repeatedly duplicate the last bit to obtain an alternating string $0101\dots$. Then, extend any run (except for the initial $0$) by duplicating the last bit of the preceding run, to obtain $01^{n_1}0^{n_2}1^{n_3}0^{n_4}\dots$. If the seed string $s$ contains a $0$, we are done. Otherwise, duplicate a $1$ bit from $s$ to obtain a $0$. Thus, $\Src_1(s)$ is fully expressive when $\abs{\Sigma}=2$.

For the other direction, assume $\abs{\Sigma}\geq 4$. Denote the letters of the seed string by $s = s_0 s_1\dots s_{n-1}$, with $s_i\in\Sigma$. Since the size of the duplication window is $k=1$, it follows that,
\[D^*(s) = \set{u_0 u_1\dots u_{n-1} ; u_i \in D^*(s_i)}.\]
Furthermore, for all $i$,
\[ D^*(s_i) \subseteq \set*{s_i,\com{s_i}}^*,\]
namely, the strings derived from the letter $s_i$ may contain only the letters $s_i$ and $\com{s_i}$. Since $\abs{\Sigma} \geq 4$, There exist $a,b \in \Sigma$ such that $\set{a,\com{a}}\neq \set{b,\com{b}}$. We contend that the string $v = (ab)^{n+1}$ is not a factor of any string in $D^*(s)$. Assume to the contrary that $v$ is a factor of some string $w \in D^*(s)$. Since $\abs{s} = n$, by the pigeonhole principal, it follows that $ab$ or $ba$ is a factor of $u_i\in D^*(s_i)\subseteq\set{s_i,\com{s_i}}^*$, for some $0\leq i \leq n-1$. This contradicts the fact that $\set{a,\com{a}}\neq \set{b,\com{b}}$. Hence, $\Src_1(s)$ is not fully expressive when $\abs{\Sigma}\geq 4$.
\end{IEEEproof}

\section{Capacity and Entropy Rate}
\label{sec:capentropy}

In this section we study the capacity and entropy rate of the reverse-complement string-duplication system. Loosely speaking, while the capacity focuses on what is possible~\cite{FarSchBru16,JaiFarBru17}, the entropy rate looks at what is probable~\cite{EliFarSchBru19}. Thus, the capacity is an upper bound on the entropy rate (see~\cite{EliFarSchBru19}).

Both the capacity and the entropy rate are hard to find exactly, and in previous works it was found exactly only for a select few cases. We mention that the case of $k=1$ was studied in~\cite{EliFarSchBru19} for a binary alphabet and a seed string $s=0$. Trivially, when $k=1$ we have full capacity, $\ccap(\Src_1(0))=1$, and by an elaborate combinatorial counting argument it was shown there that $0.8689\leq h(\Src_1(0))\leq 0.9067$. However, the case $k=1$ is degenerate in several ways, as reversal is meaningless, and duplication windows never overlap. Thus, a first true step into the reverse-complement string-duplication system requires taking a window size of $k=2$ at least. We therefore restrict ourselves in this section to the binary alphabet $\Sigma=\Z_2$, a duplication length of $k=2$, and a seed string $s=00$. Surprisingly, we show that a huge gap exists in this case between the capacity and the entropy rate.

\subsection{The Capacity}

Our strategy for finding the exact capacity of the binary $\Src_2(00)$ differs from previous approaches to the problem in~\cite{FarSchBru16,JaiFarBru17}. We first characterize all the irreducible strings. We then show that we can derive all of them from the seed string with a constant overhead. Finally, we show that this implies full capacity, i.e., $\Src_2(00)=1$.

In what follows, let us denote the set of irreducible strings with respect to the $\Src_2$ system by $\irr$. That is,
\[ \irr\eqdef \set*{ u\in\Sigma^* ; A^*(u)=\set*{u}},\]
is the sets of strings who are their own sole ancestor.

\begin{figure}[t]
    \centering
    \begin{overpic}[scale=0.45]
    {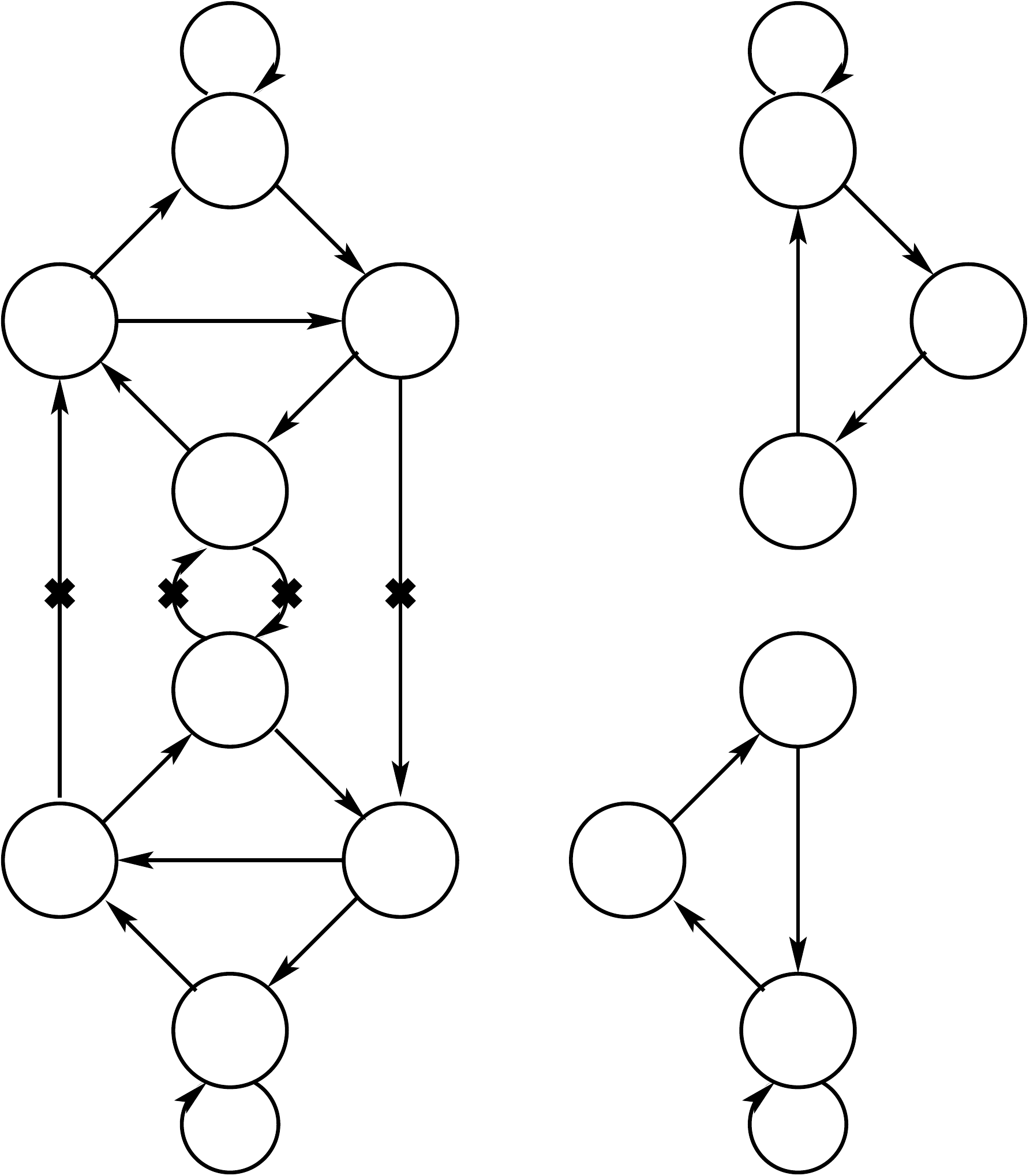}
    \put(0,95){(a)}
    \put(47,95){(b)}
    \put(17,86){$000$}
    \put(17,57){$010$}
    \put(17,40){$101$}
    \put(17,11){$111$}
    \put(2.25,71.5){$100$}
    \put(2.25,25.5){$110$}
    \put(31.5,71.5){$001$}
    \put(31.5,25.5){$011$}
    \put(25,95){$0$} 
    \put(25,4){$1$} 
    \put(26,48.5){$1$} 
    \put(11,48.5){$0$} 
    \put(36,48.5){$1$} 
    \put(1,48.5){$0$} 
    \put(28,82){$1$} 
    \put(28,63){$0$} 
    \put(28,35){$1$} 
    \put(28,16){$1$} 
    \put(9,82){$0$} 
    \put(9,63){$0$} 
    \put(9,35){$1$} 
    \put(9,16){$0$} 
    \put(19,75){$1$} 
    \put(19,22){$0$} 
    \put(73,95){$0$}
    \put(76,82){$1$}
    \put(76,63){$0$}
    \put(64,71.5){$0$}
    \put(57,35){$1$}
    \put(57,16){$0$}
    \put(73,4){$1$}
    \put(70,26){$1$}
    \end{overpic}
    \caption{The graphs used in the proof of Lemma~\ref{lemma:Irrform}: (a) The binary De Bruijn graph of order $3$, with the edges corresponding to $0011$, $0101$, $1010$, and $1100$ canceled. Labels read along paths in the graph form exactly the set of binary strings not containing the factors $0011$, $0101$, $1010$, and $1100$. (b) The minimized graph of (a).}
    \label{fig:irr}
\end{figure}
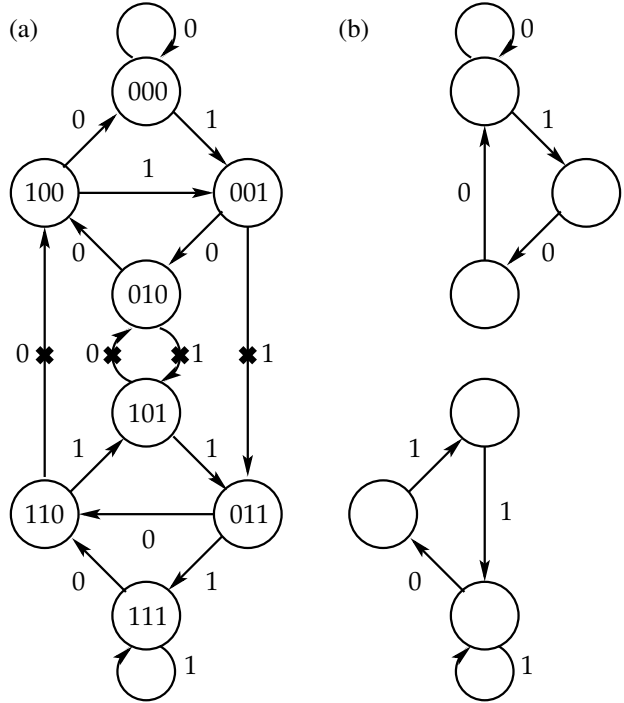

\begin{lemma}
\label{lemma:Irrform}
The set of binary irreducible strings with respect to $\Src_2$ satisfies
\[ \irr = \factor\parenv*{(0^2 0^* 1)^*}\cup \factor\parenv*{(1^2 1^* 0)^*}.\]
\end{lemma}
\begin{IEEEproof}
The irreducible strings are exactly those strings that do not contain any duplication, namely, none of the factors $0011$, $1100$, $0101$, and $1010$. Using standard techniques from constrained coding~\cite{MarRotSie01}, the set of binary strings not containing these factors is exactly generated by reading the labels on paths in the binary De Bruijn graph of order $3$, where the four edges corresponding to $0011$, $1100$, $0101$, and $1010$, are removed, seen in Figure~\ref{fig:irr}(a). By applying the Moore algorithm to minimize the number of vertices in the graph, we obtain a more compact graph representation, seen in Figure~\ref{fig:irr}(b). The set of strings generated by paths in the latter graph is immediately seen to be equal to $\factor\parenv{(0^2 0^* 1)^*}\cup \factor\parenv{(1^2 1^* 0)^*}$, as claimed.
\end{IEEEproof}

As our next step, we prove that we can derive any irreducible string from the seed string $00$ using at most a constant overhead, which is independent of the irreducible string.

\begin{lemma}
\label{lemma:reachingirr}
Denote $S=\Src_2(00)$ over $\Z_2$. Then there exists a constant $c\in\N$ such that any irreducible $u\in \irr$ is a factor of some $w\in S$ where $\abs{w}\leq\abs{u}+c$.
\end{lemma}

\begin{IEEEproof}
By Lemma~\ref{lemma:Irrform}, $u\in\irr$ is either $u\in\factor((0^2 0^* 1)^*)$ or $u\in\factor((1^2 1^* 0)^*)$. Let us first consider the former, i.e., $u\in\factor((0^2 0^* 1)^*)$. Our goal is to find a derivation $00\der^* w$ such that $u$ is a factor of $w$, with only a constant overhead.

By possibly adding a prefix and a suffix totaling no more than $4$ bits, we can find
\[ u'= 0^{\ell_1} 1 0^{\ell_2} 1 \dots 0^{\ell_m} 1,\]
$\ell_i\geq 2$ for all $i$, such that $u$ is a factor of $u'$. It then suffices to prove that we can derive $00\der^* w$ such that $u'$ is a factor of $w$. Let us further denote by $t$ the number of indices $i$ such that $\ell_i$ is even.

In preparation for finding the desired derivation, we start by deriving
\[ 00 \der^* 00 1^{2+2\ceil{t/2}} \der 001^{2+2\ceil{t/2}} 00.\]
Thus, we obtain $00 1^{2+t} 00$ if $t$ is even, and $00 1^{3+t} 00$ if $t$ is odd. We denote
\[ y \eqdef 00 1^{2+2\ceil{t/2}},  \qquad z\eqdef 00.\]
Hence, $00\der^* y z$. We shall continue by induction, at each step updating the $y$ part and the $z$, with our invariant being that the $z$ part begins with $0$ and that the $y$ part ends with sufficiently many $1$'s (which will be made clear later).

Assume now that we would like to generate a factor of $0^\ell 1$, $\ell\geq 2$, from $yz$. Further assume $y=001^i$. We distinguish between two cases, depending on the parity of $\ell$:
\begin{enumerate}
    \item
    If $\ell$ is even, assume $i\geq 3$, and then we repeatedly duplicate the penultimate $11$ in $y$ to obtain,
    \[ yz=001^{i-3}111z\der^{\ell/2} 001^{i-1} 0^\ell 1 z=y'z'.\]
    where $y'=001^{i-1}$ and $z'=0^{\ell}1z$. We consider $y'$ to be the ``new'' $y$, and similarly $z'$, and we observe that $y'$ is shorter than $y$ since a single-bit suffix of $1$ has been removed from it.
    \item
    If $\ell$ is odd, assume $i\geq 2$. We make use of the fact that $y$ ends in a $1$, and $z$ begins with a $0$, and first duplicate this $10$ factor. We then duplicate the suffix $11$ of $y$ repeatedly to obtain,
    \[ yz=001^{i-2}11z \der 001^{i-2}1101z \der^{(\ell-1)/2} 001^i 0^\ell 1 z = y'z',\]
    where $y'=y$ and $z'=0^\ell 1z$. This time, the $y$ prefix remains unchanged.
\end{enumerate}

The process described above can be repeated, namely, we can generate $0^{\ell_m}1$, then $0^{\ell_{m-1}}1$, and so on, until $0^{\ell_1}1$. This is because at each step, the $z$ part begins with a $0$, and the $y$ part ends with sufficiently many $1$'s. More precisely, each time $\ell_1$ is even, a single-bit suffix of $1$ is removed from the $y$ part, and otherwise, the $y$ part remains the same. Since we have $t$ occurrences of even $\ell_i$, and the initial $y$ part has a suffix of $2+t$ $1$'s, the process terminates with a derivation
\[ 00 \der^* 001^{2+(t\bmod 2)}0^{\ell_1}1 \dots 0^{\ell_m}100\eqdef w.\]
Thus, $w$ contains $u'$ as a factor, with at most $7$ extra bits in length. Together with the fact that $u'$ contains $u$ as a factor with at most $4$ extra bits in length we get
\[ \abs{w} \leq \abs{u}+11.\]

Finally, we need to consider the case of $u\in\factor((1^2 1^* 0)^*)$. In this case, we first derive $00\der 0011$. We ignore the $00$ prefix, and repeat the proof from the previous case, only starting with $11$ to obtain the desired derivation. In this case, we will get $2$ more extra bits of overhead, namely,
\[ \abs{w} \leq \abs{u}+13.\]
\end{IEEEproof}

We can now prove that the capacity is full.

\begin{theorem}
Over $\Z_2$,
\[ \ccap(\Src_2(00))=1.\]
\end{theorem}
\begin{IEEEproof}
Denote $S\eqdef \Src_2(00)$. We first note that all the strings in $S$ have even length. Let $c\in\N$ be the constant from Lemma~\ref{lemma:reachingirr}, and let $n\in\N$ be an even number, $n\geq c$. We contend that any $v\in\Z_2^{n-c}$ is a factor of some string in $S\cap\Z_2^n$. Indeed, let $u\in\irr$ be a root of $v$, i.e., $u$ is irreducible and $u\der^* v$. By Lemma~\ref{lemma:reachingirr}, $u$ is a factor of some string in $S\cap\Z_2^{\abs{u}+c}$. Hence, there exist $y,z\in\Z_2^*$ such that $00 \der^* yuz$ and $\abs{yz}=c$. It follows that
\[ 00 \der^* yuz \der^* yvz,\]
and $\abs{yvz}=n$.

We have reached the conclusion that all $2^{n-c}$ strings in $\Z_2^{n-c}$ appear as factors of strings in $S\cap\Z_2^n$. Since each string of length $n$ contributes at most $c+1$ distinct $(n-c)$-factors, we have
\begin{equation}
    \label{eq:factor}
\abs*{S\cap\Z_2^n} \geq \frac{2^{n-c}}{c+1}.
\end{equation}
Thus,
\[\ccap(S) = \limsup_{n\to\infty} \frac{\log_2\abs*{S\cap \Z_2^n}}{n} \geq \lim_{n \to\infty} \frac{\log_2\frac{2^{n-c}}{c+1}}{n} = 1,\]
where the inequality follows from~\eqref{eq:factor} and the fact that $\abs{S\cap\Z_2^n}=0$ for odd $n$. Since trivially $\ccap(S) \leq 1$, we have $\ccap(S) = 1$, as claimed.
\end{IEEEproof}

\subsection{Entropy Rate}

Before diving into the technical details of finding the exact entropy rate $h(\Src_2(00))$ in the binary case, we would like to describe an outline of the strategy. Recall that the object under study here is a stochastic process, starting with the seed string $00$. In each turn a position in the string is chosen, independently and uniformly, and a reverse-complement duplication of length $2$ is performed there. The process then repeats, inducing a probability distribution over outcome strings after $n$ rounds. The limit of the normalized entropy, as $n\to\infty$, gives us the entropy rate.

The approach we take is similar in spirit to~\cite{EliFarSchBru19,LouSchBruFar20}: we shall track the frequencies of $2$-factors as $n\to\infty$. Unlike previous papers, it seems to be impossible to track the frequencies exactly, and we shall have to resort to an approximate tracking only. Thus, while~\cite{EliFarSchBru19,LouSchBruFar20} manage to extract differential equations that govern the evolution of factor frequencies, we shall have to settle for a differential \emph{inclusion}. Once we have a handle on the limit of $2$-factor frequencies, we shall use semiconstrained systems~\cite{EliMeySch16} to find their exact capacity, which we show is $0$. Since the capacity upper bounds the entropy rate, we will reach the conclusion that $h(\Src_2(00))=0$.

Let us start the technical discussion. We follow~\cite{FarSchBru16,EliFarSchBru19}, and consider strings to be cyclic, namely, if $u=u_0 u_1 \dots u_{n-1}\in\Z_2^n$, then $u_{i+1}$ is the letter following $u_i$, where subscripts are taken modulo $n$. We shall find it more convenient to work with the derivative of strings rather than with the strings themselves. The \emph{derivative} of $u$ (see~\cite{Gol67}) is defined  as
\[ D(u) \eqdef u_1-u_0, u_2-u_1, \dots, u_{n-1}-u_{n-2},u_0-u_{n-1}\in \Z_2^n,\]
where the commas are written simply in order to separate the letters of the string. The derivative is easily seen to be a linear mapping which is two-to-one (mapping a string and its complement to the same derivative). 

The stochastic process we are studying, $S=\Src_2(00)$, starts with the seed string $s=00$, and we denote $S(0)=s$. For each $i=1,2,\dots$, a position in $S(i-1)$ is chosen uniformly and independently at random, and a reverse-complement duplication of length $k=2$ is performed on this position, resulting in $S(i)$. Thus, each $S(n)$, $n\in\N$, is a random variable representing the outcome of $n$ duplications. Recall that the entropy of $S(n)$ is defined as
\[ H(S(n))\eqdef -\sum_{w\in\Sigma^*}\Pr(S(n)=w)\log_2\Pr(S(n)=w),\]
where we observe that the sum is actually a finite one. With this, the \emph{entropy rate} of the random process $S$ is defined as
\[ h(S)\eqdef \limsup_{n\to\infty}\frac{1}{n}H(S(n)).\]

The probabilities $\Pr(S(n)=w)$ are hard to find, and thus, to compute $h(S)$ we resort to an indirect route. This involves statistics of factors of $S(n)$. To that end we give some more definitions. Let $u,v\in\Sigma^*$ be two strings, $\abs{u}\geq\abs{v}$. We use $\abs{u}_v$ to denote the number of times $v$ appears as a factor of $u$, including cyclically. More precisely, if the letters of $u$ and $v$ are $u=u_0 u_1 \dots u_{n-1}$ and $v=v_0 v_1 \dots v_{k-1}$, then
\[ \abs*{u}_v \eqdef \abs*{ \set*{0\leq i\leq n-1 ; u_i=v_0, u_{i+1}=v_1, \dots, u_{i+k-1}=v_{k-1}}},\]
where the indices of $u$ are taken modulo $n$. The frequency of $v$ in $u$ is then defined as
\[ \fr_{v}(u) \eqdef \frac{\abs*{u}_v}{\abs*{u}}.\]
We can then see that for cyclic strings,
\[ \fr_{v}(u) = \sum_{a\in\Sigma} \fr_{va}(u) = \sum_{a\in\Sigma} \fr_{av}(u). \]
Let us therefore define the set of $\ell$-order admissible frequency vectors as
\[ \cQ(\Sigma^\ell) \eqdef \set*{
(\alpha^v)_{v\in\Sigma^\ell}\in [0,1]^{\abs{\Sigma}^\ell}
;
\sum_{v\in\Sigma^\ell} \alpha^v =1, 
\forall v'\in\Sigma^{\ell-1}: \sum_{a\in\Sigma} \alpha^{v'a} = \sum_{a\in\Sigma} \alpha^{av'}
} .\]
Note that string indices, as the $v$ in $\alpha^v$, are written in superscript. We also comment that the $\alpha$'s in the vector $(\alpha^v)_{v\in\Sigma^\ell}$ are indexed in lexicographic order. For example, $(\alpha^{00},\alpha^{01},\alpha^{10},\alpha^{11})\in\cQ(\Z_2^2)$ if and only if $\alpha^v\in[0,1]$ for all $v\in\Z_2^2$, $\alpha^{00}+\alpha^{01}+\alpha^{10}+\alpha^{11}=1$, and also $\alpha^{00}+\alpha^{01}=\alpha^{00}+\alpha^{10}$, as well as $\alpha^{11}+\alpha^{10}=\alpha^{11}+\alpha^{01}$. We will also use partial vectors of $\cQ(\Sigma^{\ell})$ by replacing entries with a dot. For example, the statement $(\alpha^{00},\alpha^{01},\alpha^{10},\cdot)\in\cQ(\Z_2^2)$ is equivalent to stating that there exists $\alpha^{11}$ such that $(\alpha^{00},\alpha^{01},\alpha^{10},\alpha^{11})\in\cQ(\Z_2^2)$.

Our analysis of the stochastic process uses the framework of stochastic approximation~\cite{Bor08}. The fundamental tool we employ is that of the differential inclusion limit~\cite[Sec.~5]{Bor08}, and whose main theorem we now recall.

\begin{theorem}[Differential Inclusion Limit~\cite{Bor08}]
\label{th:Diffinclimit}
Let $z_n$ be a discrete stochastic process in $\R^d$ given by
\[z_{n+1} = z_n + a_n\parenv*{y_n + M_{n+1}}, \quad n\geq 0,\]
with a given $z_0$. Assume all the following conditions hold:
\begin{enumerate}
    \item[(A1)]
    $y_n \in f(z_n)$ for all $n\geq 0$, where $f$ is a set-valued map $f:\mathbb{R}^d \to \cP(\R^d)$ that satisfies:
    \begin{enumerate}
        \item[(i)] 
        For each $z \in \R^d$, $f(z)$ is convex and compact.
        \item[(ii)]
        For all $z \in \R^d$, 
        \[\sup_{y\in f(z)}\norm*{y} < K (1+\norm*{z})\]
        for some constant $K>0$.
        \item[(iii)]
        h is upper semicontinuous in the sense that if $z_n \to z$, $y_n \to y$, $y_n\in f(z_n)$, for $n\geq 1$, then $y \in f(z)$.
    \end{enumerate}
    \item[(A2)] 
    $a_n>0$ for all $n\geq 0$, are fixed positive scalars satisfying 
    \[\sum_n a_n = \infty, \qquad \sum_n a_n^2 < \infty.\]
    \item [(A3)]
    $\set{M_n}$ is a martingale difference sequence w.r.t.~the increasing
    $\sigma$-algebras $\cF_n=\sigma(z_m,y_m,M_m,m\leq n)$, $n\geq 0$, i.e.,
    \[\E\bset*{M_{n+1};\cF_n}=0,\]
    almost surely, $n\geq 0$. Additionally, $\set{M_n}$ are square-integrable with
    \[ \E\bset*{\norm*{M_{n+1}};\cF_n}\leq K\parenv*{1+\norm{z_n}^2},\]
    almost surely, $n\geq 0$, for some constant $K>0$.
    \item[(A4)]
    The sequence is bounded,
    \[\sup_n \norm{z_n}<\infty,\]
    almost surely.
\end{enumerate}
Then $z_n$ converges almost surely to a closed connected internally chain transitive invariant set of the differential inclusion limit
\begin{equation}
    \label{eq:zinc}
\frac{\mathrm{d}}{\mathrm{d}t}z(t) \in f(z(t)).
\end{equation}
\end{theorem}

The following theorem does the heavy lifting in this section, proving that the derivative of $S(n)$ is asymptotically, almost surely, composed nearly entirely of $1$'s.

\begin{theorem}
\label{th:asymfreq}
Let $S=\Src_2(00)$ be the stochastic process described above. Then, almost surely,
\[ \lim_{n\to\infty} \fr_1 \parenv*{D(S(n))} = 1.\]
\end{theorem}

\begin{IEEEproof}
While our goal is to prove a claim on the frequency of $1$'s, we shall need to track the frequencies of other factors as well. More precisely, we shall follow how the frequencies of $0$'s, $10$'s, and $11$'s evolve in $D(S(n))$ as $n\to\infty$. Let us therefore define
\[ x_n \eqdef \begin{pmatrix} \abs*{D(S(n))}_0 \\ \abs*{D(S(n))}_{10} \\ \abs*{D(S(n))}_{11} \end{pmatrix}, \qquad z_n \eqdef \begin{pmatrix} \fr_0(D(S(n))) \\ \fr_{10}(D(S(n))) \\ \fr_{11}(D(S(n))) \end{pmatrix}=\frac{1}{2n+2}x_n,\]
where the last equality is due to the fact that $\abs{S(n)}=\abs{D(S(n))}=2n+2$.

We now write
\[
x_{n+1} = x_n + \xi_{n+1},
\]
which we can rewrite in the following form:
\begin{equation}
    \label{eq:zeq}
z_{n+1} = z_n + \frac{1}{2n+4}\parenv*{\xi_{n+1}-2z_n}.
\end{equation}
The value of $\xi_{n+1}$ depends on the position of the duplication taken from $S(n)$ to $S(n+1)$. To find the possible values of $\xi_{n+1}$, consider a string $u\in\Z_2^*$ whose letters are $u_i$. If a duplication is performed on the $i$th position then 
\[u=\dots u_i u_{i+1} u_{i+2} \dots \der \dots u_i u_{i+1} \com{u_{i+1}} \com{u_i} u_{i+2} \dots.\]
In the derivative domain this becomes,
\begin{equation}
\label{eq:derchange}
\dots u_{i+1}-u_i, u_{i+2}-u_{i+1}, u_{i+3}-u_{i+2} \dots \longrightarrow \dots u_{i+1}-u_i, 1 , u_i-u_{i+1}, \com{u_{i+2}-u_i}, u_{i+3}-u_{i+2} \dots,
\end{equation}
where $\longrightarrow$ is used instead of $\Longrightarrow$ to emphasize that this is a reverse-complement duplication in the derivative domain. Tabulating all the possible cases of~\eqref{eq:derchange} gives us Table~\ref{tab:derchange}.

\begin{table}[t]
    \caption{The cases of~\eqref{eq:derchange}, showing the relevant factors of the derivative before and after the duplication, as well as the vector $\xi_{n+1}$ representing the change in the number of occurrences of the factors $0$, $10$, and $11$.}
    \label{tab:derchange}
\[
\begin{array}{r||c|c|c|c|c|c|c|c|}
\text{before dup.} & 000 & 001 & 010 & 011 & 100 & 101 & 110 & 111 \\
\hline
\text{after dup.} & 01010 & 01011 & 01000 & 01001 & 11100 & 11101 & 11110 & 11111 \\
\hline
\xi_{n+1} &
\begin{pmatrix} 0 \\ +2 \\ 0 \end{pmatrix} &
\begin{pmatrix} 0 \\ +1 \\ +1 \end{pmatrix} &
\begin{pmatrix} +2 \\ 0 \\ 0 \end{pmatrix} &
\begin{pmatrix} +2 \\ +1 \\ -1 \end{pmatrix} &
\begin{pmatrix} 0 \\ 0 \\ +2 \end{pmatrix} &
\begin{pmatrix} 0 \\ 0 \\ +2 \end{pmatrix} &
\begin{pmatrix} 0 \\ 0 \\ +2 \end{pmatrix} &
\begin{pmatrix} 0 \\ 0 \\ +2 \end{pmatrix}
\end{array}
\]
\end{table}

We further manipulate~\eqref{eq:zeq} to obtain,
\[ z_{n+1} = z_n + \frac{1}{2n+4}\parenv*{\E\bset*{\xi_{n+1};\cF_n}-2z_n+\xi_{n+1}-\E\bset*{\xi_{n+1};\cF_n}},
\]
where $\cF_n$ is the $\sigma$-algebra generated by $\sigma(z_m,\xi_m,m\leq n)$. We first observe that 
\[M_{n+1}\eqdef \xi_{n+1}-\E\bset*{\xi_{n+1};\cF_n}\]
is a martingale difference sequence w.r.t.~$\cF_n$ since
\[ \E\bset*{M_{n+1} ; \cF_n} = \E\bset*{ \xi_{n+1}-\E\bset*{\xi_{n+1};\cF_n} ; \cF_n} = \E\bset*{\xi_{n+1};\cF_n} - \E\bset*{\xi_{n+1};\cF_n} = 0.\]

Next, we study
\begin{equation}
    \label{eq:defy}
y_n \eqdef \E\bset*{\xi_{n+1};\cF_n}-2z_n.
\end{equation}
Determining $\E\bset*{\xi_{n+1};\cF_n}$ seems difficult. However, if we denote\footnote{We use superscripts here to remind us of the relevant factors, and not to represent powers.}
\begin{align*}
\alpha^{000}_n & \eqdef \E\bset*{\fr_{000}(D(S(n)));\cF_n}, & \alpha^{001}_n & \eqdef \E\bset*{\fr_{001}(D(S(n)));\cF_n}, \\
\alpha^{010}_n & \eqdef \E\bset*{\fr_{010}(D(S(n)));\cF_n}, & \alpha^{011}_n & \eqdef \E\bset*{\fr_{011}(D(S(n)));\cF_n},
\end{align*}
then by using Table~\ref{tab:derchange} we could write,
\begin{align*}
\E\bset*{\xi_{n+1};\cF_n}&=
\alpha^{000}_n \begin{pmatrix} 0 \\ +2 \\ 0 \end{pmatrix}+
\alpha^{001}_n \begin{pmatrix} 0 \\ +1 \\ +1 \end{pmatrix}+
\alpha^{010}_n \begin{pmatrix} +2 \\ 0 \\ 0 \end{pmatrix}+
\alpha^{011}_n \begin{pmatrix} +2 \\ +1 \\ -1 \end{pmatrix} \\
&\quad +(1-\alpha^{000}_n-\alpha^{001}_n-\alpha^{010}_n-\alpha^{011}_n)\begin{pmatrix} 0 \\ 0 \\ +2 \end{pmatrix} \\
&= \begin{pmatrix}
2\alpha^{010}_n+2\alpha^{011}_n \\
2\alpha^{000}_n+\alpha^{001}_n+\alpha^{011}_n \\
2-2\alpha^{000}_n-\alpha^{001}_n-2\alpha^{010}_n-3\alpha^{011}_n
\end{pmatrix}.
\end{align*}
We make the observation that, given $z_n= (z^0_n, z^{10}_n, z^{11}_n)^\T$,
\begin{equation}
\label{eq:sumfr}
\alpha^{000}_n+\alpha^{001}_n+\alpha^{010}_n+\alpha^{011}_n = z^0_n.
\end{equation}
Looking back at~\ref{eq:defy}, we can therefore write,
\begin{align*}
y_n &= \begin{pmatrix}
2\alpha^{010}_n+2\alpha^{011}_n \\
2\alpha^{000}_n+\alpha^{001}_n+\alpha^{011}_n \\
2-2\alpha^{000}_n-\alpha^{001}_n-2\alpha^{010}_n-3\alpha^{011}_n
\end{pmatrix} - 2\begin{pmatrix} z^0_n \\ z^{10}_n \\ z^{11}_n \end{pmatrix} \\
&= \begin{pmatrix}
-2 & 0 & 0 \\
0 & -2 & 0 \\
-2 & 0 & -2
\end{pmatrix}
\begin{pmatrix} z^0_n \\ z^{10}_n \\ z^{11}_n \end{pmatrix}+
\begin{pmatrix}
2\alpha^{010}_n+2\alpha^{011}_n \\
2\alpha^{000}_n+\alpha^{001}_n+\alpha^{011}_n \\
2+\alpha^{001}_n-\alpha^{011}_n
\end{pmatrix}.
\end{align*}
Let us now define the function $f:\R^3\to \cP(\R^3)$, sending vectors from $\R^3$ to subsets of $\R^3$, in the following way,
\begin{align*}
f\begin{pmatrix} z^0 \\ z^{10} \\ z^{11} \end{pmatrix} &\eqdef
\set*{
\begin{pmatrix}
-2 & 0 & 0 \\
0 & -2 & 0 \\
-2 & 0 & -2
\end{pmatrix}
\begin{pmatrix} z^0 \\ z^{10} \\ z^{11} \end{pmatrix}+
\begin{pmatrix}
2\alpha^{010}+2\alpha^{011} \\
2\alpha^{000}+\alpha^{001}+\alpha^{011} \\
2+\alpha^{001}-\alpha^{011}
\end{pmatrix}
;
(\alpha^{000},\alpha^{001},\alpha^{010},\alpha^{011},\cdot,\cdot,\cdot,\cdot)\in\cQ(\Z_2^3)
}.
\end{align*}
It then follows that $y_n$ from~\eqref{eq:defy} satisfies,
\[ y_n \in f(z_n).\]

By a simple check we can verify that the requirements of Theorem~\ref{th:Diffinclimit} are satisfied. Thus, by employing it we can say that $z_n$ converges almost surely to the limit of a function satisfying~\eqref{eq:zinc}. In our case, that means solving
\[ \frac{\mathrm{d}}{\mathrm{d}t} \begin{pmatrix} z^0 \\ z^{10} \\ z^{11} \end{pmatrix} = \begin{pmatrix}
-2 & 0 & 0 \\
0 & -2 & 0 \\
-2 & 0 & -2
\end{pmatrix}
\begin{pmatrix} z^0 \\ z^{10} \\ z^{11} \end{pmatrix}+
\begin{pmatrix}
2\alpha^{010}+2\alpha^{011} \\
2\alpha^{000}+\alpha^{001}+\alpha^{011} \\
2+\alpha^{001}-\alpha^{011}
\end{pmatrix}
\]
for some constants $(\alpha^{000},\alpha^{001},\alpha^{010},\alpha^{011},\cdot,\cdot,\cdot,\cdot)\in\cQ(\Z_2^3)$, and with the initial condition
\[ z^0(0) = 1, \qquad z^{10}(0)=0, \qquad z^{11}(0)=0,\]
since the seed string is $00$. The solution is then
\begin{align*}
z^{0}(t) &= (\alpha^{010}+\alpha^{011})(1-e^{-2t})+e^{-2t},\\
z^{10}(t) &= \parenv*{\alpha^{000}+\frac{1}{2}\alpha^{001}+\frac{1}{2}\alpha^{011}}(1-e^{-2t}),\\
z^{11}(t) &= \parenv*{1+\frac{1}{2}\alpha^{001}-\alpha^{010}-\frac{3}{2}\alpha^{011}}(1-e^{-2t})+e^{-2t}\cdot 4t(\alpha^{010}+\alpha^{011}-1).
\end{align*}
Thus, almost surely,
\begin{equation}
\label{eq:z1}
\lim_{n\to\infty}\begin{pmatrix} \fr_0(D(S(n))) \\ \fr_{10}(D(S(n))) \\ \fr_{11}(D(S(n))) \end{pmatrix}=\lim_{t\to\infty}\begin{pmatrix} z^0(t) \\ z^{10}(t) \\ z^{11}(t) \end{pmatrix} = \begin{pmatrix} \alpha^{010}+\alpha^{011} \\ \alpha^{000}+\frac{1}{2}\alpha^{001}+\frac{1}{2}\alpha^{011} \\ 1+\frac{1}{2}\alpha^{001}-\alpha^{010}-\frac{3}{2}\alpha^{011} \end{pmatrix}\eqdef \begin{pmatrix} z^0_\infty \\ z^{10}_\infty \\ z^{11}_\infty \end{pmatrix}.
\end{equation}
Since for any cyclic binary string $u$ we have
\[ \fr_0(u)+\fr_{10}(u)+\fr_{11}(u)=1,\]
this relation also holds in the limit, and therefore,
\[ z^0_\infty + z^{10}_\infty + z^{11}_\infty = 1.\]
Substituting the values from~\eqref{eq:z1} we thus have
\[ \alpha^{000}+\alpha^{001}=0.\]
Since $\alpha^{000},\alpha^{001}\in[0,1]$, we reach the conclusion that
\[ \alpha^{000}=\alpha^{001}=0,\]
and we obtain
\begin{equation}
\label{eq:z2}
\begin{pmatrix} z^0_\infty \\ z^{10}_\infty \\ z^{11}_\infty \end{pmatrix} = \begin{pmatrix} \alpha^{010}+\alpha^{011} \\ \frac{1}{2}\alpha^{011} \\ 1-\alpha^{010}-\frac{3}{2}\alpha^{011} \end{pmatrix}.
\end{equation}

The information in~\eqref{eq:z2} is not yet sufficient to prove the claim. The main technical obstacle is that Theorem~\ref{th:Diffinclimit} does not completely determine all the parameters. However, by reusing it in a slightly different manner we can improve our result. We go all the way back to our observation~\eqref{eq:sumfr}. We now make another observation, which is that the frequencies of $01$ and $10$ in a cyclic binary string must be equal. Hence,
\begin{align*}
\alpha^{010}_n+\alpha^{011}_n &=  \E\bset*{\fr_{010}(D(S(n)));\cF_n}+ \E\bset*{\fr_{011}(D(S(n)));\cF_n}\\
&= \E\bset*{\fr_{01}(D(S(n)));\cF_n}=\E\bset*{\fr_{10}(D(S(n)));\cF_n}=z^{10}_n.
\end{align*}
Continuing down the same path as before, we now have,
\begin{align*}
y_n &= \begin{pmatrix}
2\alpha^{010}_n+2\alpha^{011}_n \\
2\alpha^{000}_n+\alpha^{001}_n+\alpha^{011}_n \\
2-2\alpha^{000}_n-\alpha^{001}_n-2\alpha^{010}_n-3\alpha^{011}_n
\end{pmatrix} - 2\begin{pmatrix} z^0_n \\ z^{10}_n \\ z^{11}_n \end{pmatrix} \\
&= \begin{pmatrix}
-2 & 2 & 0 \\
1 & -2 & 0 \\
-2 & 0 & -2
\end{pmatrix}
\begin{pmatrix} z^0_n \\ z^{10}_n \\ z^{11}_n \end{pmatrix}+
\begin{pmatrix}
0 \\
\alpha^{000}_n-\alpha^{010}_n \\
2+\alpha^{001}_n-\alpha^{011}_n
\end{pmatrix},
\end{align*}
and by defining
\begin{align*}
g\begin{pmatrix} z^0 \\ z^{10} \\ z^{11} \end{pmatrix} &\eqdef
\set*{
\begin{pmatrix}
-2 & 2 & 0 \\
1 & -2 & 0 \\
-2 & 0 & -2
\end{pmatrix}
\begin{pmatrix} z^0 \\ z^{10} \\ z^{11} \end{pmatrix}+
\begin{pmatrix}
0 \\
\alpha^{000}-\alpha^{010}\\
2+\alpha^{001}-\alpha^{011}
\end{pmatrix}
;
(\alpha^{000},\alpha^{001},\alpha^{010},\alpha^{011},\cdot,\cdot,\cdot,\cdot)\in\cQ(\Z_2^3)
}.
\end{align*}
It then follows that $y_n$ from~\eqref{eq:defy} satisfies,
\[ y_n \in g(z_n).\]
Once again we use Theorem~\ref{th:Diffinclimit}, however this time we obtain that almost surely,
\begin{equation}
\label{eq:z3}
\begin{pmatrix} z^0_\infty \\ z^{10}_\infty \\ z^{11}_\infty \end{pmatrix} = \begin{pmatrix} \beta^{000}-\beta^{010} \\ \beta^{000}-\beta^{010} \\ 1-\beta^{000}+\frac{1}{2}\beta^{001}+\beta^{010}-\frac{1}{2}\beta^{011} \end{pmatrix},
\end{equation}
for some $(\beta^{000},\beta^{001},\beta^{010},\beta^{011},\cdot,\cdot,\cdot,\cdot)\in\cQ(\Z_2^3)$. We crucially note that the first two components of~\eqref{eq:z3} are equal. Carrying this knowledge back to~\eqref{eq:z2} we obtain the equation
\[\alpha^{010}+\alpha^{011}=\frac{1}{2}\alpha^{011}.\]
Since $\alpha^{010},\alpha^{011}\in[0,1]$, we necessarily have
\[\alpha^{010}=\alpha^{011}=0.\]
Putting this back into~\eqref{eq:z2}, we reach the conclusion that, almost surely,
\[ \begin{pmatrix} z^0_\infty \\ z^{10}_\infty \\ z^{11}_\infty \end{pmatrix} = \begin{pmatrix} 0 \\ 0 \\ 1 \end{pmatrix},\]
thus proving our claim.
\end{IEEEproof}

 Returning from the derivative domain is simple.

\begin{corollary}
\label{cor:asymfreq}
Let $S=\Src_2(00)$ be the stochastic process described above. Then, almost surely,
\[ \lim_{n\to\infty}
\begin{pmatrix}
\fr_{00}(S(n)) \\
\fr_{01}(S(n)) \\
\fr_{10}(S(n)) \\
\fr_{11}(S(n))
\end{pmatrix}
=\begin{pmatrix} 0 \\ \frac{1}{2} \\ \frac{1}{2} \\ 0 \end{pmatrix}.
\]
\end{corollary}
\begin{IEEEproof}
In a cyclic binary string, the frequencies of $01$ and $10$ are equal. Additionally, they are the only factors creating $1$'s in the derivative. Thus, the conclusion follows from Theorem~\ref{th:asymfreq}.
\end{IEEEproof}

\begin{figure}
    \centering
    \begin{overpic}[scale=0.6] 
    {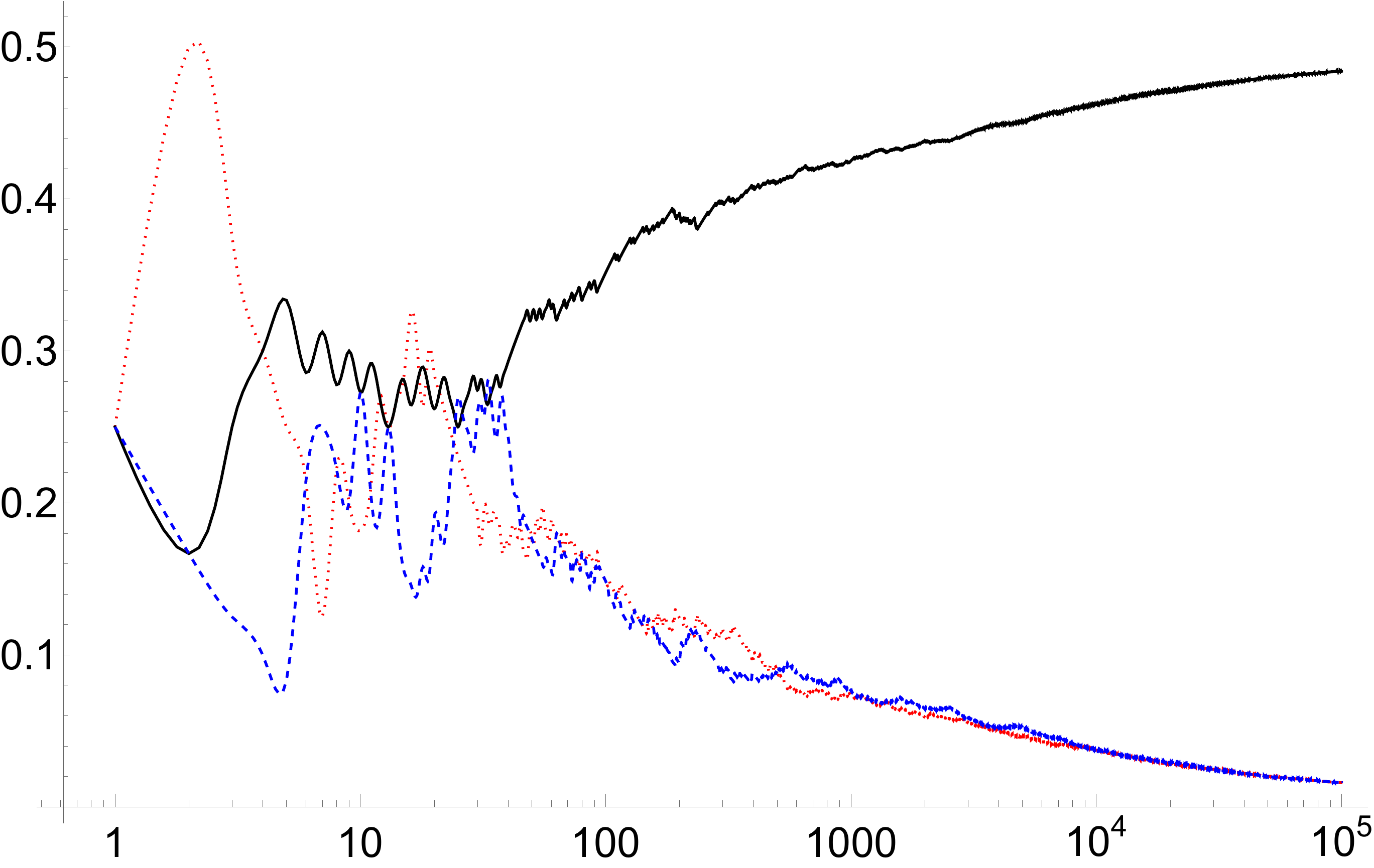}
    \put(18,55){(a)}
    \put(55,55){(b)}
    \put(22,10){(c)}
    \put(52,-2){$n$}
    \end{overpic}\\ 
    \ \\
    \caption{An example simulation of $\Src_2(00)$ showing (a) $\fr_{00}(S(n))$, (b) $\fr_{01}(S(n))=\fr_{10}(S(n))$, and (c) $\fr_{11}(S(n))$.}
    \label{fig:freq}
\end{figure}
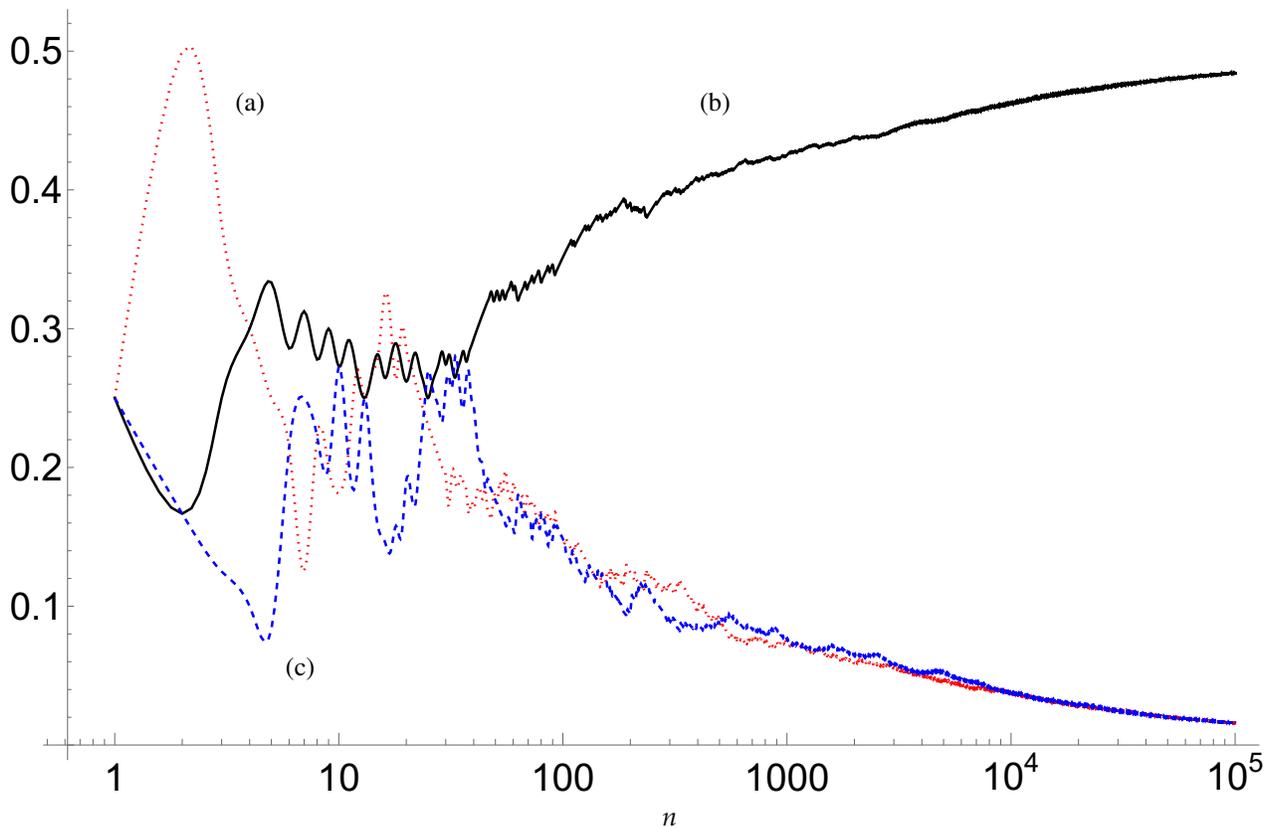

An illustration of the frequencies of $2$-factors in $S(n)$, as $n\to\infty$, is shown in Figure~\ref{fig:freq}. Notice in this figure how $\fr_{01}(S(n))=\fr_{10}(S(n))$ tends to $\frac{1}{2}$, hence in the derivative, $\fr_1(D(S(n))$ tends to $1$. We now conclude this section by proving that the entropy rate of $\Src_2(00)$ is $0$. We recall the following useful bound~\cite[Th.~11]{LouSchBruFar20}, and the remark following it.

\begin{lemma}[\cite{LouSchBruFar20}]
\label{lem:upperboundonentropy}
Let $S$ be a stochastic duplication system, as defined above. Assume that
\[ \lim_{n\to\infty} \parenv*{\fr_{v}(S(n))}_{v\in\Sigma^\ell} = (\alpha^v)_{v\in\Sigma^\ell},\]
almost surely. Then,
\[ h(S) \leq -\sum_{v\in\Sigma^\ell} \alpha^v \log_2\parenv*{\frac{\alpha^v}{\mu^v}},\]
where $\mu^v$ is the marginal on the first $\ell-1$ coordinates, namely, if $v=v_0 v_1 \dots v_{\ell-1}$, with $v_i\in\Sigma$ for all $i$, then
\[ \mu^v = \sum_{a\in\Sigma} \alpha^{v_0 \dots v_{\ell-2} a}.\]
\end{lemma}

\begin{corollary}
Over $\Z_2$,
\[ h(\Src_2(00))=0.\]
\end{corollary}
\begin{IEEEproof}
Simply combine Lemma~\ref{lem:upperboundonentropy} with Corollary~\ref{cor:asymfreq} to obtain an upper bound of $0$ on the entropy rate. A lower bound of $0$ is trivial.
\end{IEEEproof}

\section{Single-Duplication-Correcting Codes}
\label{sec:ecc}

In this section we switch gears, and consider the reverse-complement duplication as a source of noise. To protect the information against such a noise mechanism, we would like to design efficient error-correcting codes. We first formally define error-correcting codes for the reverse-complement duplication channel. We then show how to construct single-duplication-correcting codes, with odd duplication length, by altering known code constructions for single-burst-insertion correction.

\begin{definition}
An $(n,M,t)_k^\rc$ reverse-complement-duplication code is a set $C \subseteq \Sigma^n$ of size $|C| = M$, such that for every $c_1\neq c_2 \in C$, $D^t(c_1)\cap D^t(c_2) = \emptyset$, all with respect to the $\Src_k$ string-duplication system. The redundancy of the code (in bits) is defined as $\log_2(\abs{\Sigma}^n /\abs{C})$.
\end{definition}

Thus, in a code capable of correcting $t$ reverse-complement duplications of length $k$, no two distinct codewords have the same $t$-descendant. Our focus here is on single-error-correcting codes, namely, $t=1$.

Trivially a reverse-complement duplication is a special case of an insertion. Hence, a simple ``off-the-shelf'' solution to finding such a code is to employ a general insertion-correcting code. The first code solving this problem in the literature is the renowned Varshamov-Tenengolts (VT) code~\cite{VarTen65,Lev65a}. It is a binary code which addresses the case of a single insertion or deletion, i.e., $k=1$. The size of the VT code is $\approx 2^n/(n+1)$, and it has redundancy $\approx \log_2(n+1)$. This was extended to a $q$-ary alphabet in~\cite{Ten84}, with a code size $\approx q^n/(qn)$ and redundancy $\approx \log_2 n+\log_2 q$. The binary VT codes were also extended to binary codes capable of correcting (general) $k$ insertions or deletions. For example, recently, \cite{SimBru21} (see also references therein) constructed binary $k$-deletion-correcting codes with redundancy $8k\log_2 n+o(\log n)$.

Taking another step forward, we observe that a single duplication does not insert $k$ symbols in random positions, but rather as a single burst of length $k$. Thus, to correct a single duplication we may use a ready-made code capable of correcting a single burst of insertions of length $k$. A binary code that corrects a single burst of insertions or deletions of length $k=2$, was first constructed in~\cite{Lev67}. This code has size $\approx 2^n/(2n)$ and therefore redundancy $\approx 1+ \log_2 n$. Recently, ~\cite{SchWacGabYaa17} (see also references therein) constructed binary codes that correct a single burst of length (general) $k$ insertions or deletions, with redundancy $\approx \log_2 n + (k - 1)\log_2 \log_2 n + k - \log_2 k$.

Thus, in light of the above, the advantage of the construction we are about to propose is two-fold. First, except for the degenerate case of burst-length $1$ (i.e., a single inserted symbol), the known constructions are only over the binary alphabet. Our construction works for arbitrary alphabets. Second, the redundancy of our construction does not depend on the size of the alphabet.

To present our construction we require the notion of a complement-preserving mapping.

\begin{definition}
Let $\Sigma$ be any alphabet of even size, with a complement operation as defined in Section~\ref{sec:prelim}. A mapping $\beta:\Sigma\to\Z_2$ is said to be \emph{complement preserving} if for all $a\in\Sigma$,
\[ \beta(\com{a})=\com{\beta(a)}.\]
\end{definition}

Complement-preserving mappings always exist, and may be chosen to be easily computable. For example, if $\Sigma=\Z_{q}$, with $\com{2i}=2i+1$ for all $i$, then we may take $\beta(a)= a \bmod 2$. We also extend $\beta$ to act on strings in the natural way, namely, $\beta(a_0 a_1 \dots a_{n-1})\eqdef \beta(a_0)\beta(a_1)\dots\beta(a_{n-1})$, with $a_i\in\Sigma$ for all $i$. We can now give our construction.

\begin{construction}
\label{con:tobinary}
Let $C'\subseteq\Z_2^n$ be a binary code capable of correcting a single burst insertion of odd length $k$. Let $\Sigma$ be an alphabet of even size, and $\beta:\Sigma\to\Z_2$ a complement-preserving mapping. We construct the following code:
\[ C \eqdef \set*{ c\in \Sigma^n ; \beta(c)\in C'}.\]
\end{construction}

To prove that this construction is indeed capable of correcting a single reverse-complement duplication, one might expect the following straightforward decoding algorithm: upon receiving a string $w\in\Sigma^{n+k}$, the receiver computes $\beta(w)$ and then it runs the decoding algorithm for the binary burst-insertion-correcting code $C'$. The receiver then finds out what $k$-factor is to be removed from $w$ to obtain the transmitted codeword. This however is insufficient, as the following example shows. Suppose $\Sigma=\Z_4$ (with $\com{0}=1$ and $\com{2}=3$), and $k=3$. Consider the received string $w=1332202221$, for which we have $\beta(w)=1110000001$. Assume further that the decoder of $C'$ determines a $000$ factor was inserted, and is therefore to be removed. Unfortunately, there are four possible factors $000$ in $\beta(w)$, the removal of any of which results in the same string $\beta(c)=1110001$. However, in the string $w$ over $\Z_4$ these four factors are $220$, $202$, $022$ and $222$. It is not immediately clear which of them is to be removed to obtain the transmitted $c$. We solve this problem by using the fact that the inserted burst forms a reverse-complement duplication. In this case, only $220$ forms a reverse-complement duplication since it is preceded by $133$.

We recall the useful Lyndon-Sch\"utzenberger Lemma, and then state and prove the correctness of the code construction.
\begin{lemma}[{{\cite[Lemma 2]{LynSch62}}}]
\label{lem:lynsch}
Let $x,y,z\in \Sigma^*$, $x\neq \varepsilon$. If $xy=yz$, then there exist $u,v\in\Sigma^*$ and $\ell\geq 0$ such that $x=uv$, $y=(uv)^\ell u$, and $z=vu$.
\end{lemma}

\begin{theorem}
\label{th:codeproof}
The code $C$ from Construction~\ref{con:tobinary} can correct a single reverse-complement duplication of length $k$.
\end{theorem}

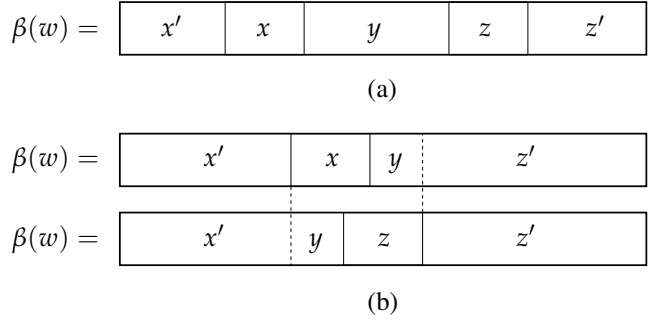
\begin{figure}[t]
    \centering
    \hspace*{9ex}
    \begin{overpic}[scale=0.35]
    {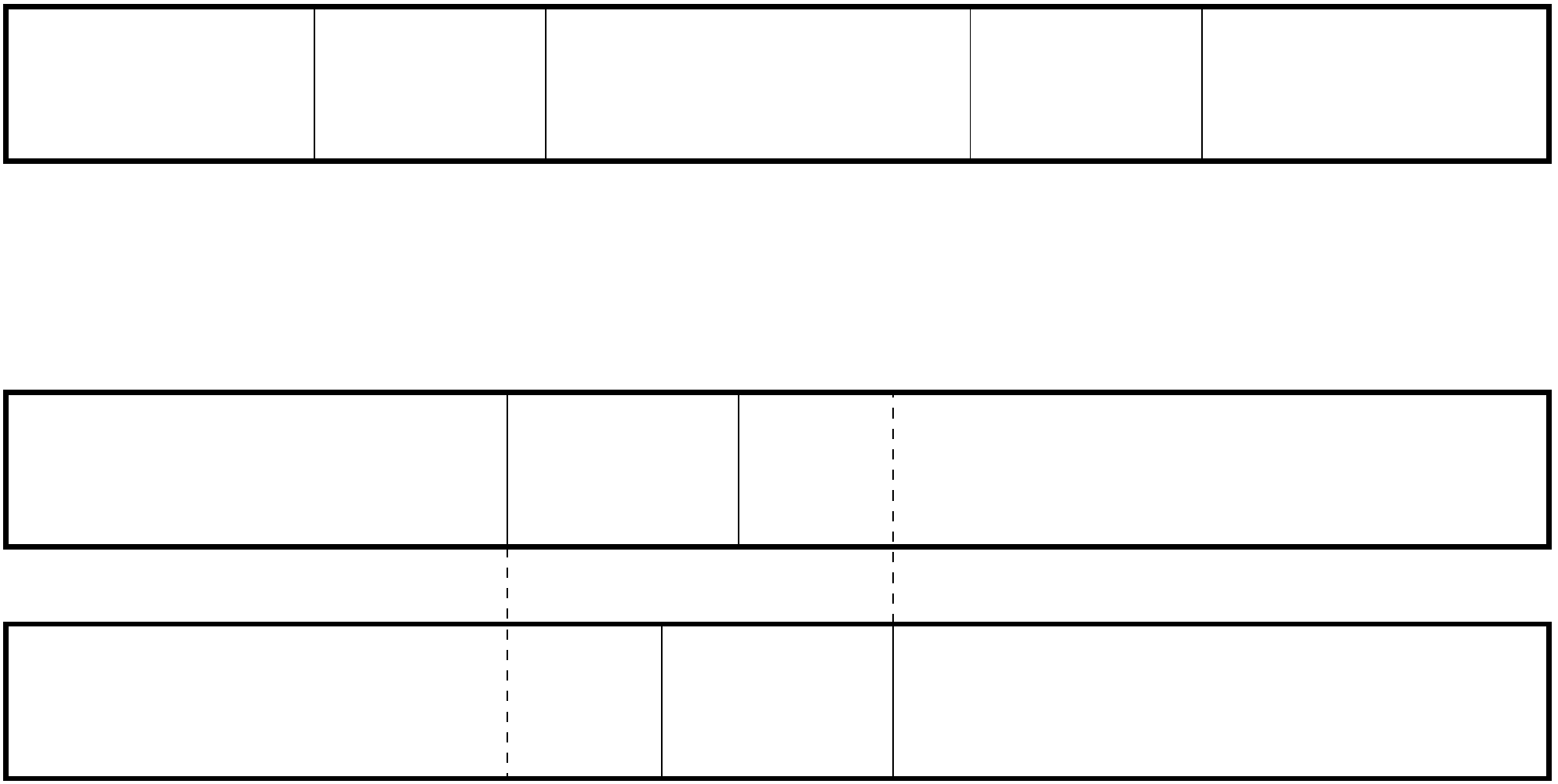}
    \put(8,43.5){$x'$}
    \put(26,43.5){$x$}
    \put(47,43.5){$y$}
    \put(68,43.5){$z$}
    \put(88,43.5){$z'$}
    \put(16,19){$x'$}
    \put(39,19){$x$}
    \put(51,19){$y$}
    \put(75,19){$z'$}
    \put(16,4){$x'$}
    \put(36,4){$y$}
    \put(49,4){$z$}
    \put(75,4){$z'$}
    \put(47,32){(a)}
    \put(47,-8){(b)}
    \put(-20,43.5){$\beta(w)=$}
    \put(-20,19){$\beta(w)=$}
    \put(-20,4){$\beta(w)=$}
    \end{overpic}
    \vspace{1em}
    \caption{The two cases of $\beta(w)$ in the proof of Theorem~\ref{th:codeproof}}
    \label{fig:dupcases}
\end{figure}

\begin{IEEEproof}
Throughout the proof, we adopt the notation that if $w\in\Sigma^n$ is a string, its letters are denoted $w_0 w_1 \dots w_{n-1}$. Assume that a codeword of $C$ was transmitted, and a single reverse-complement duplication of length $k$ occurred which
resulted in $w\in\Sigma^{n+k}$. We then compute $\beta(w)\in\Z_2^{n+k}$ and use the decoding procedure of $C'$ on it. This decoding procedure identifies at least one $k$-factor of $\beta(w)$ whose removal results in a codeword from $C'$. Assume to the contrary that two $k$-factors are identified by the decoding procedure, $x,z\in\Z_2^k$, in distinct positions. We contend that with the additional knowledge that both $x$ and $z$ form a reverse-complement duplication, a contradiction must be reached. We distinguish between two cases, depending on whether the positions of $x$ and $z$ overlap, as depicted in Figure~\ref{fig:dupcases}.

\textbf{Case I:} Assume $x$ and $z$ do not overlap (see Figure~\ref{fig:dupcases}(a)). Since the removal of $x$ and the removal of $z$ result in the same codeword of $C'$, we necessarily have $xy=yz$. By Lemma~\ref{lem:lynsch} there exist $u,v \in \Z_2^*$ and an integer $\ell \geq 0$ such that
\begin{align}
\label{eq:xyz}
    x &= uv, & y &= (uv)^\ell u, & z &= vu.
\end{align}

According to~\eqref{eq:xyz}, the $k$-factor preceding $z$ in $\beta(w)$ is $vu$. Since $z$ is a duplication of length $k$, its preceding $k$-factor must be the reverse complement of $z$. Thus, by~\eqref{eq:xyz} we get,
\[
    z= vu = (\com{vu})^R = \com{z}^R.
\]
Since $\abs{z}=k$ is odd, we have an odd-length string that is equal to its own reverse complement. The middle letter of that string, $z_i$, $i=(k-1)/2$, must therefore satisfy $z_i=\com{z_i}$, a contradiction.

\textbf{Case II:} Assume that $x$ and $z$ overlap (see Figure~\ref{fig:dupcases}(b)). In this case we must have $x=uv$, $y=u$, and $z=vu$. Since $x$ is a duplication, it is preceded by its reverse complement. Thus, the following is a factor of $\beta(w)$:
\[
\lefteqn{\overbrace{\phantom{\com{v}^R\  \com{u}^R}}^{\com{x}^R}\ \overbrace{\phantom{u\ v}}^x}
\com{v}^R\ \com{u}^R\ \,u\ \underbrace{v\ \,u}_z
\]
Since $k$ is odd, and $k=\abs{u}+\abs{v}$, we have that $\abs{u}-\abs{v}$ is odd, and in particular, $\abs{u}\neq\abs{v}$. If $\abs{u}=0$ then $x$ and $z$ are in the exact same location, which contradicts our assumption that they are not. If $\abs{v}=0$ then $x$ and $z$ do not overlap. Hence, in what follows we assume $\abs{u},\abs{v}>0$. We now use the fact that $z$ is also a duplication, and hence, must be preceded by its reverse complement. We divide our discussion depending on the relation between $\abs{u}$ and $\abs{v}$.

\begin{enumerate}
\item 
$0<\abs{v}<\abs{u}$: In order for $z$ to be a duplication the following equations must hold,
\begin{align}
v_0 v_1\dots v_{\abs{v}-1} &= \com{u_{\abs{u}-1} u_{\abs{u}-2}\dots u_{\abs{u}-\abs{v}}}, \\
u_0 u_1\dots u_{\abs{u}-\abs{v}-1} & =\com{u_{\abs{u}-\abs{v}-1} u_{\abs{u}-\abs{v}-2} \dots u_0}. \label{eq:case1.2}
\end{align}
Recall that $\abs{u}-\abs{v}$ is odd. Thus, \eqref{eq:case1.2} shows a string of odd length that equals its reverse complement. In particular, for its middle letter we get $u_i=\com{u_i}$, $i=(\abs{u}-\abs{v}-1)/2$, a contradiction.

\item $0<\abs{u}<\abs{v}<2\abs{u}$:
For $z$ to form a duplication, the following equations must hold,
\begin{align}
v_0 v_1 \dots v_{\abs{u}-1} &= \com{u_{\abs{u}-1} u_{\abs{u}-2} \dots u_0}, \label{eq:case2.1}\\
v_{\abs{u}} v_{\abs{u}+1} \dots v_{\abs{v}-1} &= u_0 u_1 \dots u_{\abs{v}-\abs{u}-1}, \\
u_0 u_1 \dots u_{2\abs{u}-\abs{v}-1} &= u_{\abs{v}-\abs{u}} u_{\abs{v}-\abs{u}+1}\dots u_{\abs{u}-1}, \\
u_{2\abs{u}-\abs{v}} u_{2\abs{u}-\abs{v}+1} \dots u_{\abs{u}-1} &= v_ 0 v_1 \dots v_{\abs{v}-\abs{u}-1}. \label{eq:case2.4}
\end{align}
From~\eqref{eq:case2.1} and~\eqref{eq:case2.4} we get that 
\[u_{2\abs{u}-\abs{v}} u_{2\abs{u}-\abs{v}+1} \dots u_{\abs{u}-1} = \com{u_{\abs{u}-1} u_{\abs{u}-2} \dots u_{2\abs{u}-\abs{v}}}.\]
Once again we have a string of odd length that is equal to its own reverse complement, a contradiction.

\item $\abs{v}=2\abs{u}$: This time for $z$ to form a duplication we must have,
\begin{align}
v_0 v_1 \dots v_{\abs{u}-1} &= \com{u_{\abs{u}-1} u_{\abs{u}-2} \dots u_0}, \label{eq:case3.1}\\
v_{\abs{u}} v_{\abs{u}+1} \dots v_{\abs{v}-1} &= u_0 u_1 \dots u_{\abs{u}-1}, \\
u_0 u_1 \dots u_{\abs{u}-1} &= v_0 v_1\dots v_{\abs{u}-1}. \label{eq:case3.3}
\end{align}
Recall that $\abs{u}-\abs{v}$ is odd, so since $\abs{v}$ is even in this case, we have that $\abs{u}$ is odd. From~\eqref{eq:case3.1} and~\eqref{eq:case3.3} it follows that $u=\com{u}^R$, which is again a contradiction since $u$ has odd length.

\item $2\abs{u}<\abs{v}$: For $z$ to form a duplication we necessarily have,
\begin{align}
v_0 v_1 \dots v_{\abs{u}-1} &= \com{u_{\abs{u}-1} u_{\abs{u}-2} \dots u_0}, \label{eq:case4.1}\\
v_{\abs{u}} v_{\abs{u}+1} \dots v_{2\abs{u}-1} &= u_0 u_1 \dots u_{\abs{u}-1}, \\
v_{2\abs{u}} v_{2\abs{u}+1} \dots v_{\abs{v}-1} &= v_0 v_1 \dots v_{\abs{v}-2\abs{u}-1}, \label{eq:case4.3}\\
u_0 u_1 \dots u_{\abs{u}-1} &= v_{\abs{v}-2\abs{u}} v_{\abs{v}-2\abs{u}+1} \dots v_{\abs{v}-\abs{u}-1}. \label{eq:case4.4}
\end{align}
By repeated applications of~\eqref{eq:case4.3} we may reduce the indices of $v$ modulo $2\abs{u}$. Denote by $I\eqdef\set{\abs{v}-2\abs{u},\abs{v}-2\abs{u}+1,\dots,\abs{v}-\abs{u}-1}$ the $\abs{u}$ consecutive indices of $v$ appearing on the right-hand side of~\eqref{eq:case4.4}. They may be reduced modulo $2\abs{u}$ to obtain $I\bmod 2\abs{u}$. We contend that the following intersection
\[ J=\parenv*{I \bmod 2\abs{u}} \cap \set*{0,1,\dots,\abs{u}-1} \]
is not empty. If this intersection were empty, then necessarily $I\bmod 2\abs{u}=\set{\abs{u},\abs{u+1},\dots,2\abs{u}-1}$. But that would mean that
\[ \abs{v}-2\abs{u} \equiv \abs{u} \pmod{2\abs{u}},\]
and therefore
\[ \abs{v}-\abs{u} \equiv 0 \pmod{2\abs{u}},\]
which implies that $\abs{v}-\abs{u}$ is even, a contradiction. Denote
\begin{align*}
s &\eqdef (\abs{v}-2\abs{u}) \bmod 2\abs{u}, \\
t &\eqdef (\abs{v}-\abs{u}) \bmod 2\abs{u}.
\end{align*}
Thus, either $J=\set{0,1,\dots,t-1}$ or $J=\set{s,s+1,\dots,\abs{u}-1}$.

Assume the former case holds, i.e., $J=\set{0,1,\dots,t-1}$. Observe that
\[ \abs{J} = t \equiv \abs{v}-\abs{u} \equiv 1 \pmod{2}.\]
Then, using only the indices in $J$ with~\eqref{eq:case4.1} and~\eqref{eq:case4.4} we have
\[ u_{\abs{u}-t} \dots u_{\abs{u}-1} = v_0 \dots v_{t-1} = \com{u_{\abs{u}-1} \dots u_{\abs{u}-t}}.\]
We thus obtained a string of odd length that equals its reverse complement, a contradiction.

If we assume the latter case holds, i.e., $J=\set{s,s+1,\dots,\abs{u}-1}$, then
\[ \abs{J} = \abs{u}-s \equiv \abs{u}-\abs{v} \equiv 1 \pmod{2}.\]
Then, using only the indices in $J$ with~\eqref{eq:case4.1} and~\eqref{eq:case4.4} we have
\[ u_{0} \dots u_{\abs{u}-s-1} = v_s \dots v_{\abs{u}-1} = \com{u_{\abs{u}-s-1} \dots u_{0}}.\]
Once again, we obtained a string of odd length that equals its reverse complement, a contradiction.
\end{enumerate}

After having considered all cases and reaching a contradiction in all of them, we are forced to deduce that no $x$ and $z$ factors are possible. That means that the decoding procedure for $C'$, when running on $\beta(w)$, may suggest several $k$-factors have been inserted, but exactly one of them forms a reverse-complement duplication.
\end{IEEEproof}

Using Theorem~\ref{th:codeproof}, we may use known binary codes that are capable of correcting a single burst insertion of odd length $k$, as the component code $C'$ in Construction~\ref{con:tobinary}. The resulting $q$-ary codes are capable of correcting a single reverse-complement duplication of length $k$. For example, taking $C'$ to be the binary VT code~\cite{VarTen65,Lev65a}, we can construct a $q$-ary $(n,M,1)_1^\rc$ code with redundancy $\approx \log_2(n+1)$ for \emph{any} even $q$. This is better than using the $q$-ary VT code~\cite{Ten84} with redundancy $\approx \log_2 n+ \log_2 q$. For odd $k\geq 3$ we can use the binary burst-insertion-correcting code of~\cite{SchWacGabYaa17} to create a $q$-ary $(n,M,1)_1^\rc$ code for \emph{any} even $q$, with redundancy $\approx \log_2 n + (k - 1)\log_2 \log_2 n + k - \log_2 k$, which does not depend on $q$.

To conclude this section, we would like to point out a peculiarity of codes that correct reverse-complement duplications. It is well known that codes that correct insertions can also correct the same amount of deletions, even when burst insertions/deletions are concerned (e.g., see~\cite{SchWacGabYaa17,Lev65a}). If insertion-correcting codes correspond to duplication-correcting codes, then deletion-correcting codes correspond to deduplication-correcting codes, which we now define.

\begin{definition}
An $(n,M,t)_k^\rc$ reverse-complement-deduplication code is a set $C \subseteq \Sigma^n$ of size $|C| = M$, such that for every $c_1\neq c_2 \in C$, $A^t(c_1)\cap A^t(c_2) = \emptyset$, all with respect to the $\Src_k$ string-duplication system.
\end{definition}

Hence, deduplication-correcting codes have non-intersecting $t$-ancestors, whereas duplication-correcting codes have non-intersecting $t$-descendants.

The peculiarity we would like to point out is that duplication-correcting codes are not necessarily deduplication-correcting codes, and vice versa. The following counter examples show this.

\begin{example}
Consider the following code,
\[C = \set*{00110,00011} \subseteq \Z_2^5.\]
This is a $(5,2,1)_2^\rc$ duplication-correcting code since,
\begin{align*}
D^1(00110) &= \set*{0011110,0010110,0011000,0011010}, \\
D^1(00011) &= \set*{0011011,0001111,0001011,0001100},
\end{align*}
hence, $D^1(00110)\cap D^1(00011) = \emptyset$. However, $C$ is \emph{not} a $(5,2,1)_2^\rc$ deduplication-correcting code since
\[ 000 \in A^1(00110)\cap A^1(00011).\]
\end{example}

\begin{example}
Consider the following code,
\[C = \set*{1100,1111} \subseteq \Z_2^4.\]
This is a $(4,2,1)_2^\rc$ deduplication-correcting code since,
\begin{align*}
A^1(1100) &= \set{11}, \\
A^1(1111) &= \emptyset,
\end{align*}
hence, $A^1(00110)\cap A^1(00011) = \emptyset$. Observe that $A^1(1111)=\emptyset$ since $1111$ is irreducible. We now also note that $C$ is \emph{not} a $(4,2,1)_2^\rc$ duplication-correcting code since 
\[110011 \in D^1(00110)\cap D^1(00011).\]
\end{example} 

\section{Conclusion}
\label{sec:conc}

In this paper, we studied the reverse-complement string-duplication system. When viewed as a generative system, we fully classified the cases in which $\Src_k(s)$ has full expressiveness. Interestingly, these differ depending on whether $k$ is even or odd, and with $k=1$ being a special case altogether. We then focused on the binary case with $k=2$ and a seed string $00$. We proved that the capacity in this case is full, i.e., $\ccap(\Src_2(00))=1$, but the entropy rate vanishes, i.e., $h(\Src_2(00))=0$. We switched gears to look at the duplications as a noise source, and constructed $q$-ary codes that correct a single reverse-complement duplication from binary burst-insertion-correcting codes. The construction works for odd duplication lengths, and produces codes whose redundancy (in bits) does not depend on the alphabet size.

We find the fact that $h(\Src_2(00))=0$ particularly surprising. When looking at previously known results~\cite{EliFarSchBru19}, we have $\ccap(\Src_1(0))=1$ and $0.8689\leq h(\Src_1(0))\leq 0.9067$. In stark contrast, when the duplication length is increased from $k=1$ to $k=2$ we still have $\ccap(\Src_2(00))=1$, but $h(\Src_2(00))=0$. To the best of our knowledge, this is the first case of a duplication system that is fully expressive, with full capacity, but vanishing entropy rate. Why that is the case when $k=2$, and whether it remains so for larger values of $k$, is still unknown. Additionally, from Figure~\ref{fig:freq} it appears as if the convergence to the limit is quite slow.

Many open questions remain, and we mention a few. First, we observe that our proof that the binary $\Src_2(00)$ has full capacity is tailored to this case, and relies on the classification of irreducible strings in $\Src_2$. The use of irreducible strings to prove full capacity is new, and we suspect that the same approach may work for any alphabet size, and any duplication length. However, the classification of irreducible strings in each case becomes increasingly complex. It would be interesting to find a more general approach to the problem of determining $\ccap(\Src_k(s))$ over any alphabet. A similar open question pertains to the problem of finding the entropy rate $h(\Src_k(s))$ for general alphabets. The stochastic approximation method we used calls for a decision on what factors to follow. It is still unknown how to generalize this to cases other than $\Src_2(00)$.

Finally, the construction for duplication-correcting codes that we provided only works for odd duplication lengths. It is interesting to find a construction for even $k$, and to find out whether it offers the same savings in redundancy compared with burst-insertion-correcting codes. In addition, the construction only works for a single duplication, which parallels the fact that only single-burst-insertion-correcting codes are known. These problems, and others, are left for future work.

\bibliographystyle{IEEEtranS}
\bibliography{allbib}

\begin{thebibliography}{10}
\providecommand{\url}[1]{#1}
\csname url@samestyle\endcsname
\providecommand{\newblock}{\relax}
\providecommand{\bibinfo}[2]{#2}
\providecommand{\BIBentrySTDinterwordspacing}{\spaceskip=0pt\relax}
\providecommand{\BIBentryALTinterwordstretchfactor}{4}
\providecommand{\BIBentryALTinterwordspacing}{\spaceskip=\fontdimen2\font plus
\BIBentryALTinterwordstretchfactor\fontdimen3\font minus
  \fontdimen4\font\relax}
\providecommand{\BIBforeignlanguage}[2]{{%
\expandafter\ifx\csname l@#1\endcsname\relax
\typeout{** WARNING: IEEEtranS.bst: No hyphenation pattern has been}%
\typeout{** loaded for the language `#1'. Using the pattern for}%
\typeout{** the default language instead.}%
\else
\language=\csname l@#1\endcsname
\fi
#2}}
\providecommand{\BIBdecl}{\relax}
\BIBdecl

\bibitem{nimbus}
\BIBentryALTinterwordspacing
 [Online]. Available:
  \url{https://nimbusdata.com/docs/ExaDrive-DC-Datasheet.pdf}
\BIBentrySTDinterwordspacing

\bibitem{AloBruFarJai17}
N.~Alon, J.~Bruck, F.~Farnoud, and S.~Jain, ``Duplication distance to the root
  for binary sequences,'' \emph{IEEE Trans.~Inform.~Theory}, vol.~63, no.~12,
  pp. 7793--7803, Dec. 2017.

\bibitem{Bal13}
F.~Balado, ``Capacity of {DNA} data embedding under substitution mutations,''
  \emph{IEEE Trans.~Inform.~Theory}, vol.~59, no.~2, pp. 928--941, Feb. 2013.

\bibitem{Bor08}
V.~Borkar, \emph{Stochastic Approximation -- A Dynamical System
  Viewpoint}.\hskip 1em plus 0.5em minus 0.4em\relax Cambridge University
  Press, 2008.

\bibitem{ButGilSte02}
D.~K. Butler, D.~Gillespie, and B.~Steele, ``Formation of large palindromic
  {DNA} by homologous recombination of short inverted repeat sequences in
  {S}accharomyces cerevisiae,'' \emph{Genetics}, vol. 161, no.~3, pp.
  1065--1075, 2002.

\bibitem{CheChrKiaNgu19}
Y.~M. Chee, J.~Chrisnata, H.~M. Kiah, and T.~T. Nguyen, ``Deciding the
  confusability of words under tandem repeats in linear time,'' \emph{ACM
  Trans. on Algorithms}, vol.~15, no.~3, pp. 42:1--42:22, 2019.

\bibitem{CleRisBan99}
C.~T. Clelland, V.~Risca, and C.~Bancroft, ``Hiding messages in {DNA}
  microdots,'' \emph{Nature}, vol. 399, no. 6736, pp. 533--534, 06 1999.

\bibitem{DieTanBerYaoTap10}
S.~J. Diede, H.~Tanaka, D.~A. Bergstrom, M.-C. Yao, and S.~J. Tapscott,
  ``Genome-wide analysis of palindrome formation,'' \emph{Nature Genetics},
  vol.~42, no.~4, pp. 279--279, 2010.

\bibitem{EliFarSchBru19}
O.~Elischo, F.~Farnoud, M.~Schwartz, and J.~Bruck, ``The entropy rate of some
  {P}ólya string models,'' \emph{IEEE Trans.~Inform.~Theory}, vol.~65, no.~12,
  pp. 8180--8193, Dec. 2019.

\bibitem{EliMeySch16}
O.~Elishco, T.~Meyerovitch, and M.~Schwartz, ``Semiconstrained systems,''
  \emph{IEEE Trans.~Inform.~Theory}, vol.~62, no.~4, pp. 811--824, Apr. 2016.

\bibitem{ErlZie17}
Y.~Erlich and D.~Zielinski, ``{DNA Fountain} enables a robust and efficient
  storage architecture,'' \emph{Science}, vol. 355, no. 6328, pp. 950--954,
  2017.

\bibitem{FarSchBru16}
F.~Farnoud, M.~Schwartz, and J.~Bruck, ``The capacity of string-duplication
  systems,'' \emph{IEEE Trans.~Inform.~Theory}, vol.~62, no.~2, pp. 811--824,
  Feb. 2016.

\bibitem{Gol67}
S.~W. Golomb, \emph{Shift Register Sequences}.\hskip 1em plus 0.5em minus
  0.4em\relax Holden-Day, San Francisco, 1967.

\bibitem{HeiBar07}
D.~Heider and A.~Barnekow, ``{DNA}-based watermarks using the {DNA-Crypt}
  algorithm,'' \emph{BMC Bioinformatics}, vol.~8, no.~1, pp. 1--10, 2007.

\bibitem{JaiFarBru17}
S.~Jain, F.~Farnoud, and J.~Bruck, ``Capacity and expressiveness of genomic
  tandem duplication,'' \emph{IEEE Trans.~Inform.~Theory}, vol.~63, no.~10, pp.
  6129--6138, Oct. 2017.

\bibitem{JaiFarSchBru17a}
S.~Jain, F.~Farnoud, M.~Schwartz, and J.~Bruck, ``Duplication-correcting codes
  for data storage in the {DNA} of living organisms,'' \emph{IEEE
  Trans.~Inform.~Theory}, vol.~63, no.~8, pp. 4996--5010, Aug. 2017.

\bibitem{Kov19}
M.~Kova{\v{c}}evi{\'c}, ``Zero-error capacity of duplication channels,''
  \emph{IEEE Trans.~Communications}, vol.~67, no.~10, pp. 6735--6742, 2019.

\bibitem{KovTan18}
M.~Kova{\v{c}}evi{\'c} and V.~Y. Tan, ``Asymptotically optimal codes correcting
  fixed-length duplication errors in {DNA} storage systems,'' \emph{IEEE
  Comm.~Letters}, vol.~22, no.~11, pp. 2194--2197, 2018.

\bibitem{Lanetal01}
E.~S. Lander, L.~M. Linton, B.~Birren, C.~Nusbaum, M.~C. Zody, J.~Baldwin,
  K.~Devon, K.~Dewar, M.~Doyle, W.~FitzHugh \emph{et~al.}, ``Initial sequencing
  and analysis of the human genome,'' \emph{Nature}, vol. 409, no. 6822, pp.
  860--921, 2001.

\bibitem{LenWacYaa19}
A.~Lenz, A.~Wachter-Zeh, and E.~Yaakobi, ``Duplication-correcting codes,''
  \emph{Designs, Codes and Cryptography}, vol.~87, pp. 277--298, 2019.

\bibitem{Lev65a}
V.~I. Levenshtein, ``Binary codes capable of correcting deletions, insertions
  and reversals (in {R}ussian),'' \emph{Doklady Akademii Nauk SSSR}, vol. 163,
  no.~4, pp. 845--848, 1965, {E}nglish translation in {\em Soviet Physics
  Dokl.}, 10(8):707--710, 1966.

\bibitem{Lev67}
------, ``Asymptotically optimum binary code with correction for losses of one
  or two adjacent bits (in {R}ussian),'' \emph{Problemy Kibernetiki}, vol.~19,
  pp. 293--298, 1967, {E}nglish translation in {\em Systems Theory Research},
  19:293--298, 1970.

\bibitem{Lev87}
G.~Levinson and G.~A. Gutman, ``Slipped-strand mispairing: a major mechanism
  for {DNA} sequence evolution.'' \emph{Molecular Biology and Evolution},
  vol.~4, no.~3, pp. 203--221, 1987.

\bibitem{Micetal12}
M.~Liss, D.~Daubert, K.~Brunner, K.~Kliche, U.~Hammes, A.~Leiherer, and
  R.~Wagner, ``Embedding permanent watermarks in synthetic genes,'' \emph{PLoS
  ONE}, vol.~7, no.~8, p. e42465, 08 2012.

\bibitem{LouSchBruFar20}
H.~Lou, M.~Schwartz, J.~Bruck, and F.~Farnoud, ``Evolution of {$k$}-mer
  frequencies and entropy in duplication and substitution mutation systems,''
  \emph{IEEE Trans.~Inform.~Theory}, vol.~66, no.~5, pp. 3171--3186, 2020.

\bibitem{LynSch62}
R.~C. Lyndon and M.~P. Sch{\"u}tzenberger, ``The equation {$a^M=b^N c^P$} in a
  free group,'' \emph{Michigan Math.~J.}, vol.~9, pp. 289--298, 1962.

\bibitem{MarRotSie01}
\BIBentryALTinterwordspacing
B.~H. Marcus, R.~M. Roth, and P.~H. Siegel, ``An introduction to coding for
  constrained systems,'' Oct. 2001, unpublished Lecture Notes. [Online].
  Available: \url{www.math.ubc.ca/~marcus/Handbook}
\BIBentrySTDinterwordspacing

\bibitem{SchWacGabYaa17}
C.~Schoeny, A.~Wachter-Zeh, R.~Gabrys, and E.~Yaakobi, ``Codes correcting a
  burst of deletions or insertions,'' \emph{IEEE Trans.~Inform.~Theory},
  vol.~63, no.~4, pp. 1971--1985, Apr. 2017.

\bibitem{ShiNivMacChu17}
S.~L. Shipman, J.~Nivala, J.~D. Macklis, and G.~M. Church, ``{CRISPR-Cas}
  encoding of digital movie into the genomes of a population of living
  bacteria,'' \emph{Nature}, vol. 547, pp. 345--349, Jul. 2017.

\bibitem{SimBru21}
J.~Sima and J.~Bruck, ``On optimal {$k$}-deletion correcting codes,''
  \emph{IEEE Trans.~Inform.~Theory}, vol.~67, no.~6, pp. 3360--3375, Jun. 2021.

\bibitem{Ger08}
G.~R. Smith, ``Meeting {DNA} palindromes head-to-head,'' \emph{Genes Dev},
  vol.~22, pp. 2612--2620, 2008.

\bibitem{TanFar21}
Y.~Tang and F.~Farnoud, ``Error-correcting codes for noisy duplication
  channels,'' \emph{IEEE Trans.~Inform.~Theory}, no.~6, pp. 3452--3463, Jun.
  2021.

\bibitem{TanFar22}
------, ``Error-correcting codes for short tandem duplication and edit
  errors,'' \emph{IEEE Trans.~Inform.~Theory}, 2022, to appear.

\bibitem{TanYehSchFar20}
Y.~Tang, Y.~Yehezkeally, M.~Schwartz, and F.~Farnoud, ``Single-error detection
  and correction for duplication and substitution channels,'' \emph{IEEE
  Trans.~Inform.~Theory}, vol.~66, no.~11, pp. 6908--6919, Nov. 2020.

\bibitem{Ten84}
G.~Tenengolts, ``Nonbinary codes, correcting single deletion or insertion,''
  \emph{IEEE Trans.~Inform.~Theory}, vol.~30, no.~5, pp. 766--769, Sep. 1984.

\bibitem{VarTen65}
R.~R. Varshamov and G.~M. Tenengolts, ``Code which correct single asymmetric
  errors (in {R}ussian),'' \emph{Avtomatika i Telemekhanika}, vol.~26, no.~2,
  pp. 288--292, 1965, {E}nglish translation in {\em Automation and Remote
  Control}, 26(2):286--290, 1965.

\bibitem{WonWonFoo03}
P.~C. Wong, K.-k. Wong, and H.~Foote, ``Organic data memory using the {DNA}
  approach,'' \emph{Commun. ACM}, vol.~46, no.~1, pp. 95--98, Jan. 2003.

\bibitem{YehSch20}
Y.~Yehezkeally and M.~Schwartz, ``Reconstruction codes for {DNA} sequences with
  uniform tandem-duplication errors,'' \emph{IEEE Trans.~Inform.~Theory},
  vol.~66, no.~5, pp. 2658--2668, May 2020.

\bibitem{YehSch21}
------, ``Uncertainty of reconstruction with list-decoding from
  uniform-tandem-duplication noise,'' \emph{IEEE Trans.~Inform.~Theory},
  vol.~67, no.~7, pp. 4276--4287, Jul. 2021.

\bibitem{ZerEsmGul19}
M.~Zeraatpisheh, M.~Esmaeili, and T.~A. Gulliver, ``Construction of tandem
  duplication correcting codes,'' \emph{IET Communications}, vol.~13, no.~15,
  pp. 2217--2225, 2019.

\end{thebibliography}

\end{document}